\documentclass[useAMS,usenatbib]{mn2e}

\usepackage{times}
\usepackage{epsfig}
\usepackage{textcomp}
\title[On the nature of unabsorbed Seyfert 2 galaxies]{On the nature of unabsorbed Seyfert 2 galaxies}

\author[M. Brightman and K. Nandra]{Murray Brightman$\thanks{E-mail: m.brightman06@imperial.ac.uk}$ and Kirpal Nandra\\
Astrophysics Group, Imperial College London, Blackett Laboratory, Prince Consort Road, London SW7 2AW\\}

\begin{document}

\date{Accepted 2008 August 14. Received 2008 August 13; in original form 2008 July 30}

\pagerange{\pageref{firstpage}--\pageref{lastpage}} \pubyear{2008}

\maketitle

\label{firstpage}

\begin{abstract}
We present an analysis of six 12 ${\umu}$m selected Seyfert 2 galaxies that have been reported to be unabsorbed in the X-ray.  By comparing the luminosities of these galaxies in the mid-IR (12 ${\umu}$m), optical ([O\,{\sc iii}]) and hard X-ray (2-10 keV), we show that they are all under-luminous in the 2-10 keV X-ray band. Four of the objects exhibit X-ray spectra indicative of a hard excess, consistent with a heavily obscured X-ray component and hence a hidden nucleus.  In these objects the softer X-rays may be dominated by a strong soft scattered continuum or contamination from the host galaxy, which is responsible for the unabsorbed X-ray spectra observed, and accounts for the anomalously low 2-10 keV X-ray luminosity. We confirm this assertion in NGC 4501 with a {\it Chandra} observation, which shows hard X-ray emission coincident with the nucleus, consistent with heavy absorption, and a number of contaminating softer sources which account for the bulk of the softer emission. We point out that such ``Compton thick" sources need not necessarily present iron K$\alpha$ emission of high equivalent width. An example in our sample is IRASF 01475-0740, which we know must host an obscured AGN as it hosts a hidden broad line region seen in scattered light \citep{tran03}. The X-ray spectrum is nonetheless relatively unobscured and the iron K$\alpha$ line only moderate in strength ($\sim 160$ eV). These observations can be reconciled if the hidden nuclear emission is dominated by transmitted, rather than reflected X-rays, which can then be weak compared to the soft scattered light or galactic emission even at 6.4 keV.  Despite these considerations, we conclude that two sources, NGC 3147 and NGC 3660, may intrinsically lack a broad line region (BLR), confirming the recent results of \cite{bianchi08} in the case of NGC 3147.  Neither X-ray spectrum shows signs of hidden hard emission  and both sources exhibit X-ray variability leading us to believe we are viewing the nucleus directly. 
\end{abstract}


\section{Introduction}
The current model of AGN unification (e.g. \cite{antonucci93}) states that the absence of broad lines in the optical spectra of Seyfert 2 galaxies is due to their obscuration by an optically thick structure.  Following this, the obscuration that blocks the BLR from view should also absorb soft X-rays from the continuum source in the same region as the BLR.  This is well represented and documented in the prototypical Seyfert 2, NGC 1068, which shows an X-ray reflection spectrum and strong iron K$\alpha$ emission \citep{iwasawa97}, typical of Compton thick obscuration of the X-ray source, as well as displaying polarised broad lines (PBLs) in its optical spectrum, indicative of a hidden BLR \citep{antonucci85}. The X-ray line of sight column density ($N_{\rm{H}}$) has been measured to be in excess of $10^{23}$ cm$^{-2}$ in many Seyfert 2s \citep{awaki91}, greatly exceeding the typical values in Seyfert 1s and lending strong support to the unification model. However, there remains an enigmatic group of Seyfert 2s that appear unabsorbed in X-rays (e.g. \cite{pappa01}; \cite{panessa02}), whose properties therefore seem to  contradict the unified scheme. This apparent mismatch may result from a genuine absence of the BLR. This intriguing possibility has been proposed to pertain in Seyfert 2 galaxies with particularly low accretion rates, with concomitant implications for the origin of the broad optical lines \citep{nicastro03}. Apparent confirmation of such a BLR-free AGN has recently been presented by Bianchi et al. (2008) who considered the case of NGC 3147. They were able to reject an alternative interpretation that the mismatch between X-ray and optical properties was due to variability. They also argued against a hidden nucleus based on the {\it XMM-Newton} spectrum of NGC 3147, which shows a relatively modest iron K$\alpha$ emission line. Most objects with very heavily obscured nuclei show very strong iron K$\alpha$ lines \citep{turner97} with the most extreme examples being the Compton thick sources, with lines often in excess of 1 keV equivalent width \citep{matt96_2}. 

On the other hand, it is conceivable that in these unabsorbed sources we are simply being mislead by the X-ray data. The emission from the nucleus could be contaminated by diffuse or extra-nuclear point-source emission from the host galaxy or by AGN X-rays scattered by hot electrons.  It is therefore possible that there are many Seyfert 2s which from their X-ray spectra, appear to be unabsorbed, but in fact are hiding a deeply buried AGN and that the unabsorbed profile is not necessarily nuclear. The observed equivalent width of the iron K$\alpha$ line is strongly dependent on the relative contributions of the various components and, in addition, the geometry and physical parameters of the obscuring material.  In this paper we present an analysis of six apparently unabsorbed Seyfert 2 galaxies, aiming to probe the nature of these objects.   

\begin{table*}
 \centering
 \begin{minipage}{140mm}
 \caption{Positional and observational data on the six Seyfert 2 galaxies featured.  Spectropolarimetry data are taken from Tran (2003)  and positional, redshift and distance data are taken from NED.}\label{obsdat}
  \begin{tabular}{|l l l l l l l c l l l}
\hline
Object name 	& RA 		& Dec 		& z 	& Distance 	& $N_{\rm{H}}$ 		& PBLs? & $\frac{\rm{[N\,II]}}{\rm{H\alpha}}$ & $\frac{\rm{[O\,III]}}{\rm{H\beta}}$ & Ref. \\ 
(1) 		& (2) 		& (3) 		& (4) 	& (5) 		& (6) 			& (7) 	& (8) 	& (9) 	& (10)\\ 
\hline
IRASF 01475-0740& 01h50m02.7s 	& -07d25m48s 	& 0.0177& 69.7 		& $2.0 \times 10^{20}$ 	& yes 	& 0.49 	& 5.21 	& 2 \\
NGC 3147 	& 10h16m53.6s 	& +73d24m03s 	& 0.0094& 39.8 		& $2.9 \times 10^{20}$ 	& no  	& 2.71 	& 6.11 	& 1 \\
NGC 3486 	& 11h00m23.9s 	& +28d58m29s 	& 0.0023& 13.5		& $1.9 \times 10^{20}$ 	& -  	& 1.05 	& 4.57 	& 1 \\
NGC 3660 	& 11h23m32.3s 	& -08d39m31s 	& 0.0123& 56.2		& $3.4 \times 10^{20}$ 	& no 	& 0.82 	& 2.63 	& 3 \\
NGC 3976	& 11h55m57.6s 	& +06d45m03s 	& 0.0083& 39.4		& $1.1 \times 10^{20}$ 	& -  	& 1.96 	& 3.52 	& 1 \\
NGC 4501	& 12h31m59.2s 	& +14d25m14s 	& 0.0076& 36.0		& $2.6 \times 10^{20}$ 	& no 	& 2.10 	& 5.25 	& 1 \\
\hline
\end{tabular}
Col. (1) Galaxy name as given by \cite{rush93} in the extended 12 micron galaxy sample; Col. (2) Right ascension (NED, J2000); Col. (3) Declination (NED, J2000); Col. (4) Redshift (NED); Col. (5) Luminosity distance (NED, Mpc); Col. (6) Galactic $N_{\rm{H}}$ (NED, cm$^{-2}$); Col. (7) Indicates whether the objects have polarised broad lines in the optical spectrum if data exists \citep{tran03}; Cols. (8-10) Line ratios as measured by (1) \cite{ho97}; (2) \cite{degrijp92}; (3) \cite{gu06}.
\end{minipage}
\end{table*}

\section[]{Data Analysis}

\subsection[]{Sample Selection}
The parent sample for our study is the extended {\it IRAS} 12 $\umu$m sample of \cite{rush93}. From this we select objects defined as Seyfert 2s  by the NASA/IPAC Extragalactic Database (NED). We perform our own post-hoc analysis of the optical spectra below.  We require there to be good quality X-ray data ({\it XMM-Newton, Chandra or ASCA}) with a column density indicating that they are unabsorbed in X-rays ($N_{\rm{H}} < 10^{22}$ cm$^{-2}$), from the literature or our own analysis.  The NED optical classifications may not necessarily be robust, so we searched the literature to deselect sources with incorrect or ambiguous optical classification. The remaining six objects, IRASF 01475-0740, NGC 3147, NGC 3486, NGC 3660, NGC 3976, NGC 4501 form our sample: they are all unambiguously classified as  AGN using line ratio diagnostics (Line ratios and references given in Table \ref{obsdat} and BPT diagram presented in Fig. \ref{bptdiag}).

\subsection[]{Optical data}
\begin{figure}
 \includegraphics[width=67mm,angle=90]{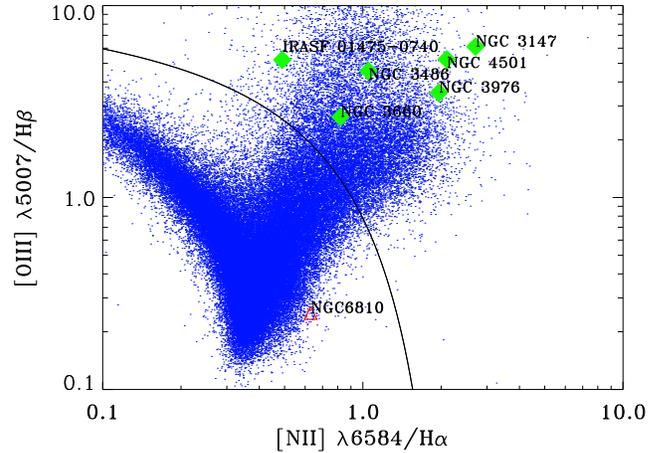}
 \caption{BPT diagram (Baldwin et al. 1981) of the six featured Seyfert 2 (green diamonds) sources plus NGC 6810 (red triangle), which has previously but incorrectly been classified as a Seyfert 2 (Strickland 2007), plotted on top of 55, 757 low redshift SDSS galaxies (blue dots) of Kauffmann
et al. (2003). The demarcation line is that defined by Kewley et al. (2001) and separates AGN from starbursts.}\label{bptdiag}
\end{figure}

We compiled optical spectroscopy, spectropolarimetry and line ratio data for our objects from the literature and present these data with references in Table \ref{obsdat} and Fig. \ref{bptdiag}, which plots the ratio [O\,{\sc iii}] $\lambda5007$/H$\beta$ versus the ratio [N\,{\sc ii}] $\lambda6584$/H$\alpha$.  We plot these on top of the catalogue of low redshift SDSS galaxies of \cite{kauffmann03}. The demarcation line is that defined by \cite{kewley01} separating AGN and starbursts, hence showing that their AGN classification is not in doubt.  Additionally, we plot the line ratios of NGC 6810, a Seyfert 2 also unabsorbed in X-rays, but shown to have a dubious Seyfert 2 classification due to broader-than-normal optical lines produced by a super-wind \citep{strickland07}.  The authors used an $XMM-Newton$ observation to show that the X-ray emission from this galaxy was probably due to X-ray binaries. 

\subsection[]{X-ray Data}
We carried out the X-ray analysis on {\it XMM-Newton} observations of IRASF 01470-0740, NGC 3147, NGC 3486, NGC 3976 and NGC 4501; {\it Chandra} observations of NGC 3147 and NGC 4501 and an {\it ASCA} observation of NGC 3660 (X-ray observational information listed in table \ref{xobsdat}).  For {\it XMM-Newton} data we use {\sc sas v7.0} tasks to perform spectral extractions on the EPIC-pn data, for the {\it Chandra} ACIS-S data we use {\sc ciao v3.4} tasks and use {\it ASCA} data products from the {\it Tartarus} database.  All spectra were grouped with a minimum of 20 counts per bin, with the exception of the {\it Chandra} observation of NGC 4501, where the spectrum was grouped using a minimum of 7 counts.  Spectral fitting was carried out using {\sc xspec v11.3}.  

\begin{table}
 \centering
   \caption{Information on the X-ray observations used in this analysis.  Exposure time in brackets give the time after filtering for background flares in {\it XMM-Newton} data.}\label{xobsdat}
  \begin{tabular}{l l l l l}
\hline
Object name	& Observatory		& Date		& Obs. ID		& exp.		\\ 
		&			&		&			& (ks)		\\
\hline
F01475-0740 	& {\it XMM-Newton}	& 2004-02-01	& 0200431101		& 12(9) 	\\
NGC 3147 	& {\it XMM-Newton}	& 2006-10-06	& 0405020601		& 17(14)	\\
		& {\it Chandra}		& 2001-09-19	& 1615			& 2		\\
NGC 3486 	& {\it XMM-Newton} 	& 2001-05-09	& 0112550101		& 15(4)  	\\
NGC 3660 	& {\it ASCA}		& 1995-06-09	& 73039000		& 26 		\\
NGC 3976	& {\it XMM-Newton}	& 2006-06-16	& 0301651801		& 14(4) 	\\
NGC 4501	& {\it XMM-Newton} 	& 2001-12-04	& 0112550801		& 14(3) 	\\ 
		& {\it Chandra}		& 2002-12-09	& 2922 			& 18 		\\ 
\hline
\end{tabular}
\end{table}

\section[]{Estimates of Bolometric Luminosity}
\begin{figure}
 \includegraphics[width=67mm,angle=90]{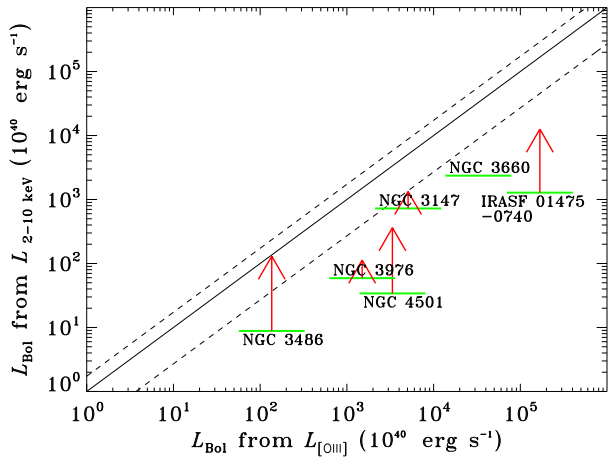}
 \caption{Estimates of the bolometric luminosities from $L_{{\rm [O\,III]}}$ with $1\sigma$ variance (green lines) against the estimated bolometric luminosities from $L_{\rm{HX}}$.  The solid black line displays where on the plot these estimates would lie should they be equal, as expected, and the dashed black lines represent the $1\sigma$ spread in $\kappa_{\rm HX}$.  As these objects are all positioned under the solid line, they are all under-luminous in X-rays.  Red arrows show how the bolometric luminosity predicted from $L_{\rm{HX}}$ changes when we try to estimate the intrinsic $L_{\rm{HX}}$ of the nucleus based on the X-ray spectral fitting.}\label{lumfig}
% \label{fig:xrb}
\end{figure}

We estimate the bolometric luminosities, $L_{\rm{Bol}}$, of the six AGN from their 12 ${\umu}$m and [O\,{\sc iii}] fluxes using bolometric corrections, $\kappa_{\rm 12\umu m}$ calculated from template SEDs of \cite{mrr08} and $\kappa_{\rm [O\,III]}$ published by \cite{heckman04}.  We also estimate bolometric luminosities from the unabsorbed 2-10 keV (HX) luminosity using the mean bolometric correction, $\kappa_{\rm HX}$, of \cite{vasudevan07}.  Table \ref{lumins} presents the observed luminosities with these estimates and Fig. \ref{lumfig} plots the estimated bolometric luminosities from the [O\,{\sc iii}] luminosity against the estimated bolometric luminosities from the 2-10 keV luminosity.  From this analysis, all six objects appear to be significantly under-luminous in the 2-10 keV X-ray band by factors of 10-100.  For a typical Seyfert 2 this is easily understood, as we expect the 2-10 keV X-rays to be suppressed by absorption, but of course for our sample it seems the evidence for that absorption in terms of the spectral shape is absent. As discussed below, despite the lack of any obvious soft X-ray absorption in these objects, some of the X-ray spectra present evidence for hidden hard components. This allows an additional estimate of the intrinsic AGN bolometric luminosity show as the red arrows in Fig. \ref{lumfig}. These are discussed in more detail below. 

\begin{table*}
 \centering
 \begin{minipage}{140mm}
  \caption{Observed luminosities of the sample using distances from Table \ref{obsdat}. $12 \umu$m luminosity is calculated from the {\it IRAS} flux density presented by Rush et al. 1993 and [O\,{\sc iii}] luminosities have been corrected for absorption using the formulation of Bassani et al. (1999). The respective predicted bolometric luminosity calculated from these measured luminosities are also given.}\label{lumins}
  \begin{tabular}{|l | c c c|c c c|}
\hline
Object name & \multicolumn{3}{|c|}{Observed luminosities (erg s$^{-1}$)} & \multicolumn{3}{|c|}{$L_{\rm{Bol}}$ (erg s$^{-1}$) predicted from:}\\
 &$L_{\rm{12{\umu}m}}$ & $L_{\rm [O\,III]}$ & $L_{\rm{HX}}$ & $L_{\rm{12{\umu}m}}$ & $L_{\rm [O\,III]}$ & $L_{\rm{HX}}$\\  
\hline
IRASF 01475-0740& $7.68\times 10^{43}$ & $4.80\times 10^{41}$ & $4.71\times 10^{41}$ & $8.54\times 10^{44}$ & $1.68\times 10^{45}$ & $1.28\times 10^{43}$\\
NGC 3147 	& $5.02\times 10^{43}$ & $1.50\times 10^{40}$ & $2.68\times 10^{41}$ & $5.58\times 10^{44}$ & $5.08\times 10^{43}$ & $7.25\times 10^{42}$\\
NGC 3486 	& $2.83\times 10^{42}$ & $3.89\times 10^{38}$ & $3.26\times 10^{39}$ & $3.14\times 10^{43}$ & $1.36\times 10^{42}$ & $8.82\times 10^{40}$\\
NGC 3660 	& $3.15\times 10^{43}$ & $9.38\times 10^{40}$ & $8.72\times 10^{41}$ & $3.50\times 10^{44}$ & $3.28\times 10^{44}$ & $2.36\times 10^{43}$\\
NGC 3976 	& $1.59\times 10^{43}$ & $4.30\times 10^{39}$ & $2.18\times 10^{40}$ & $1.76\times 10^{44}$ & $1.50\times 10^{43}$ & $5.90\times 10^{41}$\\
NGC 4501 	& $6.79\times 10^{43}$ & $9.57\times 10^{39}$ & $1.26\times 10^{40}$ & $7.54\times 10^{44}$ & $3.35\times 10^{43}$ & $3.42\times 10^{41}$\\

\hline
\end{tabular}
\end{minipage}
\end{table*}

\section[]{X-Ray Analysis}
\subsection{IRASF 01475-0740}
The {\it XMM-Newton} spectrum of IRASF 01475-0740 is well fitted by a power-law absorbed by a column of $N_{\rm{H}} = 4.1 \times 10^{21}$ cm$^{-2}$.  While this will produce some reddening and extinction in the optical, it is unlikely to be sufficient to suppress the optical broad lines entirely unless the absorber has an anomalously high dust-to-gas ratio (c.f. MCG-6-30-15; \cite{reynolds97}).

The X-ray spectrum also shows line emission between 6 and 7 keV.  The centroid energy of this line is not well constrained, so we assume that the emission is from neutral iron K$\alpha$ at a rest energy of 6.4 keV, though it may originate from ionised iron.  A gaussian fit to this feature with the centroid energy fixed at 6.4 keV gives an equivalent width (EW) of 160 eV. On the face of it, this relatively modest  iron K$\alpha$ EW argues against a hidden Compton thick nucleus (Bianchi et al. 2008). We found, however, that it was possible to fit a second, heavily absorbed component from new Monte-Carlo models of  X-rays in heavily obscured AGN, incorporating line emission (Brightman et al., in preparation), similar to those presented by e.g. \cite{ghisellini94} and \cite{krolik94}. For simplicity given the limitations of the data we fit the extreme case of 4$\pi$ coverage (i.e. a spherical distribution of matter), and we constrained the column density of the heavily absorbed component to $N_{\rm{H}}=2 \times 10^{24}|^{+65}_{-0.8}$ cm$^{-2}$. Although it is not formally required in the fit, this demonstrates a clear scenario in which the multiwavelength data can be reconciled, as such a column is easily sufficient to suppress the direct nuclear emission in the optical. By correcting for absorption in this heavily obscured component, we can make another prediction of the intrinsic luminosity of the nucleus and calculate the corresponding bolometric luminosity. This now agrees well with estimations from $L_{\rm [O\,III]}$ and $L_{\rm{12{\umu}m}}$.  The {\it XMM-Newton} observation also shows no clear signs of variability. The spectropolarimetric survey of \cite{tran03} reveals that this AGN hosts a HBLR, in full agreement with our conclusion of a heavily buried nuclear contribution. The same electrons which scatter the broad optical lines into the line of sight can then also scatter nuclear X-rays accounting for the apparently unobscured nature of the soft spectrum.  

\subsection{NGC 3147}
Our analysis of the {\it XMM-Newton} spectrum of NGC 3147 shows no absorption above the Galactic column, however it features a small (130 eV) iron K$\alpha$ line at 6.4 keV. These results are in full agreement with those of Bianchi et al. (2008).  \cite{bianchi08} used the small EW of the iron K$\alpha$ line to argue against the Compton thick nature of this source and came to the conclusion that NGC 3147 has an intrinsic absence of a BLR.  However, as we have shown in IRASF01475-0740, this conclusion is not necessarily robust. Fitting the Monte Carlo models to the spectrum once again shows that a very heavily obscured component is consistent with the spectrum. Here the column density is constrained to be $N_{\rm{H}}=9 \times 10^{23}|^{+419}_{-1.0}$ cm$^{-2}$, with the main constraint coming from the iron K$\alpha$ line.  We can again use this second component to predict the intrinsic luminosity of this source and calculate the corresponding bolometric luminosity, which also agrees well with estimations from $L_{\rm [O\,III]}$ and $L_{\rm{12{\umu}m}}$.  Other evidence suggests, however, that this may not be the correct interpretation in this case. The 2ks {\it Chandra} observation of NGC 3147 shows that the measured flux of the nucleus is $\sim2$ times that measured by {\it XMM-Newton} indicating that it is variable in nature and hence it must be being observed directly, rather than in scattered light. Furthermore,  the {\it Chandra} imaging reveals no evidence for extra-nuclear X-ray sources that could be producing the unabsorbed X-ray profile or the variability, and spectral extraction from the nuclear region in the {\it Chandra} image does not reveal a significantly harder X-ray spectrum then the larger {\it XMM-Newton} beam (Table~4). 

\subsection{NGC 3486}
The {\it XMM-Newton} spectrum of NGC 3486 is well fitted by a power-law absorbed by the Galactic column only and so appears to be another unabsorbed Seyfert 2 galaxy.  However, an excess at hard energies points to a different scenario. The spectrum is also well fitted by the addition of a Compton reflection model ({\tt pexmon}, \cite{nandra07}) with an underlying thermal component ({\tt raymond}), so this could in fact be a Compton thick Seyfert 2 which looks unabsorbed.  We also attempt to add a strongly absorbed transmission component to the model, which also improves the fit from a simple power-law.  However, the column density of this component is poorly constrained, so we fix this to an arbitrary log($N_{\rm{H}}$) = 24.5. Estimates of the intrinsic $L_{\rm{HX}}$ from the reflection component predict a $L_{\rm{Bol}}$ which agrees well with $L_{\rm{Bol}}$ estimated from $L_{\rm [O\,III]}$, suggesting that NGC 3486 is under-luminous in X-rays due to Compton thick obscuration. Finally, there is no evidence for variability of the X-ray flux in the {\it XMM-Newton} observation.

\subsection{NGC 3660}
\begin{figure}
 \centering
 \includegraphics[width=84mm]{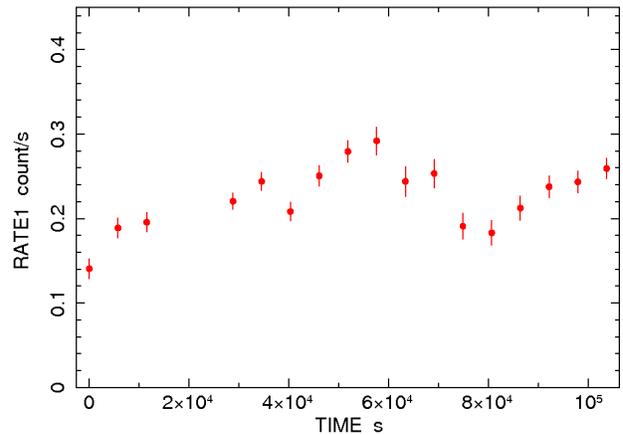}
 \caption{{\it ASCA} lightcurve of NGC 3660 showing the X-ray variable nature of the source.}\label{ngc3660lc}
% \label{fig:xrb}
\end{figure}

As there have been no observations of NGC 3660 with {\it XMM-Newton} or {\it Chandra}, we use the {\it ASCA} observation and its documentation in the {\it Tartarus} catalogue.  The {\it ASCA} spectrum is fitted well by a power-law with no absorption above the Galactic column. Fig. \ref{ngc3660lc} shows the light-curve for the 26 ks observation which shows significant variability on short time-scales. 

\subsection{NGC 3976}
The {\it XMM-Newton} spectrum of NGC 3976 is very similar to that of NGC 3486 as it shows no intrinsic absorption, but a hard excess present above the simple power-law fit allows us to add a heavily absorbed power-law component. A strongly absorbed transmission component produces an overall better fit, suggesting that NGC 3976 is also hiding a heavily obscured nucleus beneath its apparently unabsorbed soft X-ray spectrum. Again, there is also no variability detected in this observation.

\subsection{NGC 4501}
\begin{figure}
 \centering
 \includegraphics[width=50mm]{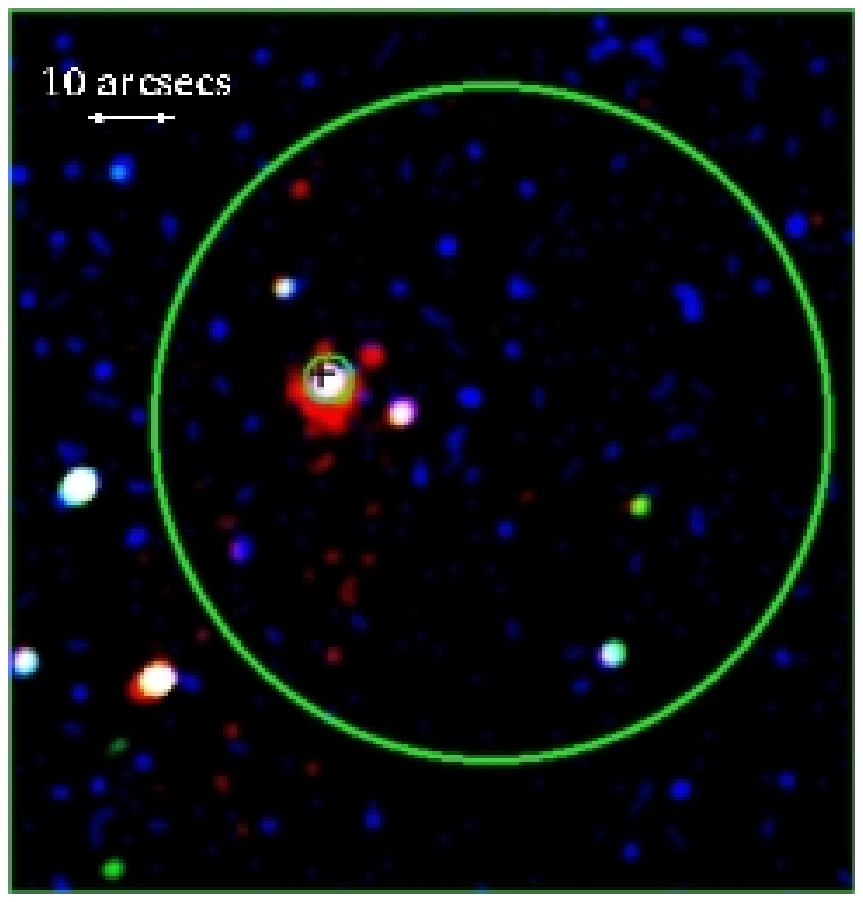}
 \caption{Chandra image of NGC 4501 smoothed by a $\sigma=3$ pixels Gaussian. Red represents 0.5-1.0 keV emission, green represents 1.0-3.0 keV emission and blue represents 3.0 - 10.0 keV in a log scale.  The black cross marks the position of the optical nucleus, the smaller green circle marks the {\it Chandra} extraction region and larger green circle marks the equivalent {\it XMM-Newton} extraction region.  The white area coincident with the nucleus indicates that it is the source of hard X-ray emission, and therefore likely to be heavily absorbed.}\label{chanimg}
% \label{fig:xrb}
\end{figure}

The {\it XMM-Newton} spectrum of NGC 4501 reveals another apparently unabsorbed Seyfert 2 galaxy in X-rays.  It is well fitted by a power-law with no intrinsic absorption plus emission from a thermal plasma component, but shows no clear hard excess or other spectral evidence supporting a deeply buried AGN (such as iron K$\alpha$ emission).  An entirely different picture emerges, however, when one considers the {\it Chandra} data, which have higher spatial resolution. The {\it Chandra} image of NGC 4501 reveals a hard X-ray emission coincident with the optically defined nucleus (Fig. \ref{chanimg}) consistent with heavy X-ray absorption. It also shows that there are also multiple extra-nuclear sources present, including diffuse soft emission close to the nucleus, which {\it XMM-Newton} could not resolve. The implication is that the unabsorbed nature of the {\it XMM-Newton} spectrum, which has a larger beam size, is due to contamination, and that the true nuclear emission is heavily obscured. We performed a spectral extraction of the optically defined nucleus using the region identified in Fig \ref{chanimg} which did indeed reveal a hard excess above the unabsorbed power-law, as in NGC 3486 and NGC 3976, which we fit with a Compton reflection component ({\tt pexmon}). Using the reflection component to estimate the intrinsic $L_{\rm{HX}}$ shows that the X-ray faintness of NGC 4501 is probably due to heavy absorption (Fig. \ref{lumfig}).  There is also no evidence for variability of NGC 4501, either between subsequent observations by {\it XMM-Newton} and {\it Chandra}, or during them.

\begin{table*}
 \centering
 \begin{minipage}{140mm}
  \caption{X-ray spectral fitting data for the sample, where R is the ratio of the underlying transmitted or reflected component to the power-law}\label{fitdat}
  \begin{tabular}{l l l l l l l l l l r@{}l}
\hline
Object name 	& Model	& {$N_{\rm{H,1}}$} 	& {$N_{\rm{H,2}}$} 	& $\Gamma$ 		& EW$_{6.4}$ 	& kT 			& R			& $\chi_r^2/\nu$& {$F_{\rm HX}$}& \multicolumn{2}{c}{$L_{\rm int}$}\\ 
(1)		& (2)	& (3)  			& (4)	 		& (5)  			& (6) 		& (7)			& (8)			& (9)		& (10) 		& & (11)	\\
\hline
IRASF01475-0740 & A	& $0.41^{+0.07}_{-0.05}$& -			& $2.02^{+0.06}_{-0.06}$& $160^{+300}_{-160}$& -		& -			& 65.5/58	& 0.84		& 49 &		\\
({\it XMM-Newton})& C	& $0.41^{+2.14}_{-0.18}$& -			& $2.15^{+0.45}_{-0.24}$& - 		& - 			& $1.0^{+32}_{-1.0}$	& 64.4/56 	& 0.84		& 48 &		\\
		& D	& $0.46^{+0.17}_{-0.17}$& $200^{+6500}_{-80}$	& $2.16^{+0.26}_{-0.23}$& - 		& - 			& $10^{+1600}_{-10}$	& 63.3/56	& 0.82		& 465 &		\\
\hline
NGC 3147 	& A	& $0.03^{+0.01}_{-0.01}$& -			& $1.60^{+0.06}_{-0.06}$& $130^{+80}_{-100}$& -			& -			& 103/113	& 1.4 		& 27 &		\\
({\it XMM-Newton})& C	& $0.03^{+0.04}_{-0.01}$& -			& $1.63^{+0.09}_{-0.07}$& - 		& -    			& $1.6^{+8.3}_{-1.6}$	& 103/112	& 1.4 		& 42 &		\\
\vspace{0.1in}	& D	& $0.04^{+0.01}_{-0.02}$& $90^{+4210}_{-10}$	& $1.68^{+0.04}_{-0.11}$& -		& -    			& $1.9^{+10}_{-1.9}$	& 102/112	& 1.4 		& 49 &		\\
NGC 3147	& A	& $<0.07$		& -			& $1.50^{+0.19}_{-0.12}$& -		& -			&  -			& 62.2/49	& 3.2		& 61 &		\\
({\it Chandra})	&	&			& 			&			&		&			&			&		&		& &		\\ 
\hline
NGC 3486 	& A	& $0.03^{+0.17}_{-0.03}$& -			& $2.44^{+1.17}_{-0.61}$& - 		& - 			& -			& 28.3/18	& 0.04		& 0 & .10	\\
({\it XMM-Newton})& B	& $<0.10$		& -			& $1.73^{+1.29}_{-0.83}$& -		& $0.23^{+0.13}_{-0.09}$& -			& 24.8/16	& 0.09		& 0 & .20	\\
		& C	& $<0.06$		& -			& $1.9^{\dag}$		& -		& $0.24^{+0.19}_{-0.12}$& $34^{+220}_{-34}$	& 21.5/15	& 0.15		& 4 & .8	\\
		& D	& $<0.06$		& 320$^{\dag}$		& $1.9^{\dag}$		& -		& $0.24^{+0.66}_{-0.10}$& $200^{+650}_{-200}$	& 23.2/16	& 0.09		& 31 & 		\\

\hline
NGC 3660 	& A	& $0.03^{\dag}$		& -			& $1.82^{+0.04}_{-0.04}$& - 		& -    			& -			& 301/304	& 2.3		& 87 &		\\
({\it ASCA})	&	&			& 			&			&		&			&			&		&		& &		\\
\hline
NGC 3976	& A	& $0.06^{+0.15}_{-0.06}$& -			& $1.80^{+1.15}_{-0.60}$& - 		& -    			& -			& 38.0/33	& 0.07		& 1 & .3	\\
({\it XMM-Newton})& B	& $0.10^{+0.10}_{-0.07}$& -			& $1.9^{\dag}$		& -		& $0.44^{+0.29}_{-0.18}$& -			& 26.2/32	& 0.05		& 0 & .97	\\
 		& C 	& $0.09^{+0.09}_{-0.06}$& -			& $1.9^{\dag}$		& - 		& $0.43^{+0.83}_{-0.14}$& $43^{+49}_{-43}$	& 27.0/30  	& 0.10		& 37 &		\\
		& D	& $0.10^{+0.06}_{-0.04}$& $17^{+190}_{-12}$	& $1.9^{\dag}$		& -		& $0.44^{+0.18}_{-0.11}$& $3.0^{+25}_{-3.0}$	& 24.8/33	& 0.12		& 4 & .2	\\

\hline
NGC 4501	& A 	& $0.21^{+0.33}_{-0.14}$& -			& $3.91^{+2.76}_{-1.14}$& - 		& - 			& -			& 21.2/13	& 0.02		& 0 & .36	\\
\vspace{0.1in}
({\it XMM-Newton})& B	& $0.02^{+0.10}_{-0.02}$& -			& $2.13^{+1.14}_{-0.63}$& -		& $0.36^{+0.20}_{-0.08}$& -			& 6.82/11	& 0.08		& 1 & .2	\\
NGC 4501	& A	& $<0.02$		& -			& $1.49^{+0.29}_{-0.36}$& -		& -			& -			& $44.7^C$	& 0.05		& 0 & .78	\\
({\it Chandra})	& B	& $1.75^{+0.12}_{-0.31}$& -			& $1.65^{+0.25}_{-0.31}$& -		& $0.05^{+0.01}_{-0.01}$& -			& $22.7^C$	& 0.07		& 1 & .1	\\
		& C 	& $0.66^{+1.11}_{-0.66}$& -			& $1.9^{\dag}$		& -		& $0.39^{+0.16}_{-0.10}$& $29^{+32}_{-20}$	& $8.64^C$	& 0.08		& 13 &		\\
		& D 	& $0.16^{+0.47}_{-0.16}$& $18^{+47}_{-15}$	& $1.9^{\dag}$		& -		& $0.35^{+0.34}_{-0.17}$& $8.9^{+28}_{-6.4}$	& $10.7^C$	& 0.09		& 2 & .6	\\

\hline
\end{tabular}
Notes: HX = 2-10 keV; $^\dag$ indicates a fixed parameter; $^C$ indicates use of Cash statistics in spectral fitting; Col. (1) Galaxy name as given by \cite{rush93}; Col. (2) Model used for fitting, A=pl, B=pl+therm, C=pl+therm+refl, D=pl+therm+trans; where pl = simple power-law; trans = Monte-Carlo model of transmitted component; therm = thermal component (raymond); refl = pure reflection component (pexmon); Col. (3) Column density of simple power-law, units of $10^{22}$ cm$^{-2}$; Col. (4) Column density of transmitted component, units of $10^{22}$ cm$^{-2}$; Col. (5) Power-law index, pegged for simple power-law and transmitted component; Col. (6) Equivalent width of the neutral iron line at 6.4 keV if present in eV; Col. (7) Temperature of the thermal component in keV; Col. (8) The normalisation ratio of the reflection or transmitted components to the simple power-law; Col. (9) Reduced chi-squared and number of degrees of freedom; Col. (10) Absorption corrected observed flux in the hard (2-10 keV) band, units of $10^{-12}$ erg cm$^{-2}$ s$^{-1}$; Col. (11) Hard X-ray intrinsic luminosity, units of $10^{40}$ erg s$^{-1}$, calculated from reflection or transmitted component where present.
\end{minipage}
\end{table*}

\begin{figure*}
\begin{minipage}{160mm}
 \includegraphics[width=80mm]{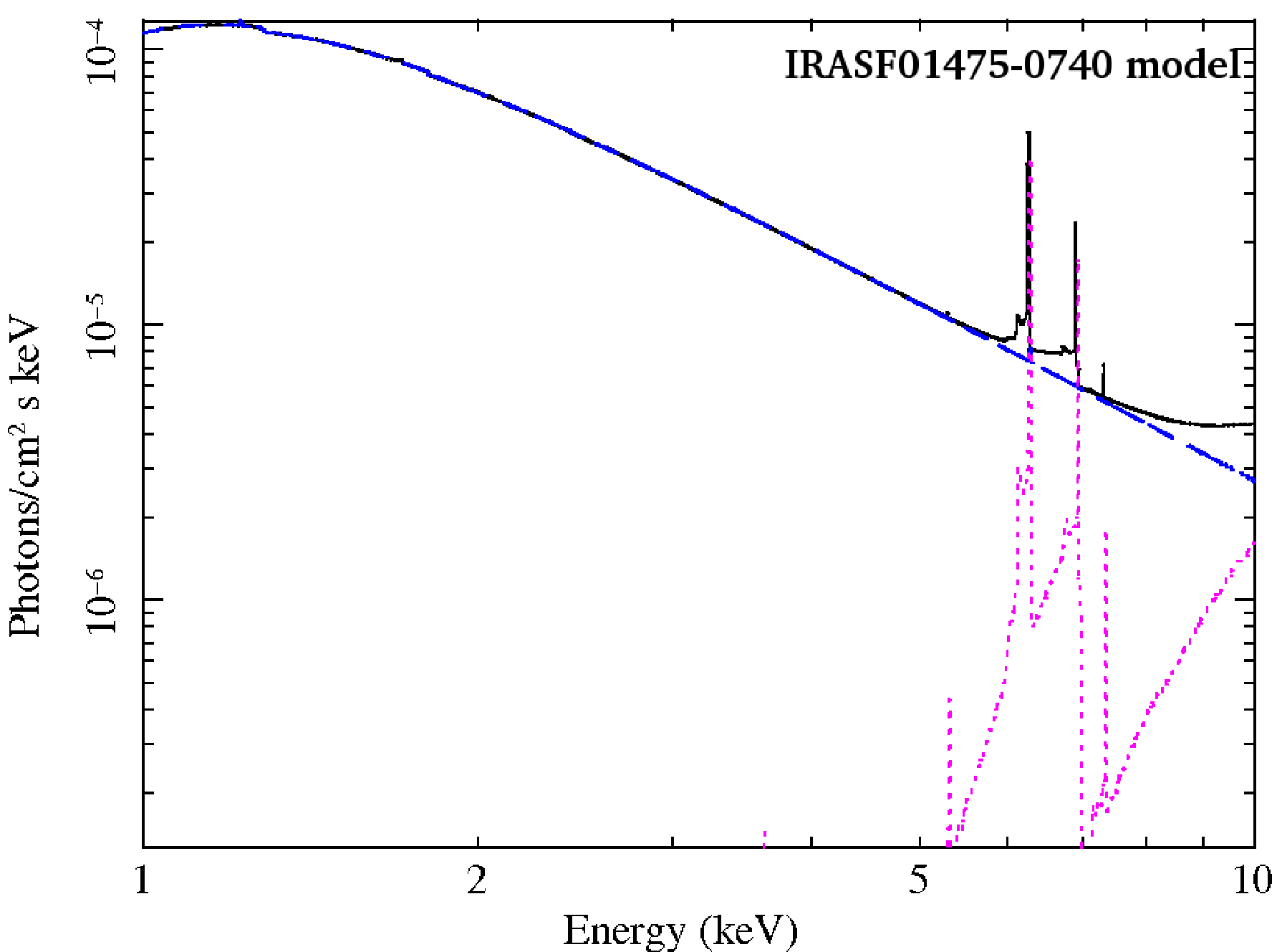}
 \includegraphics[width=80mm]{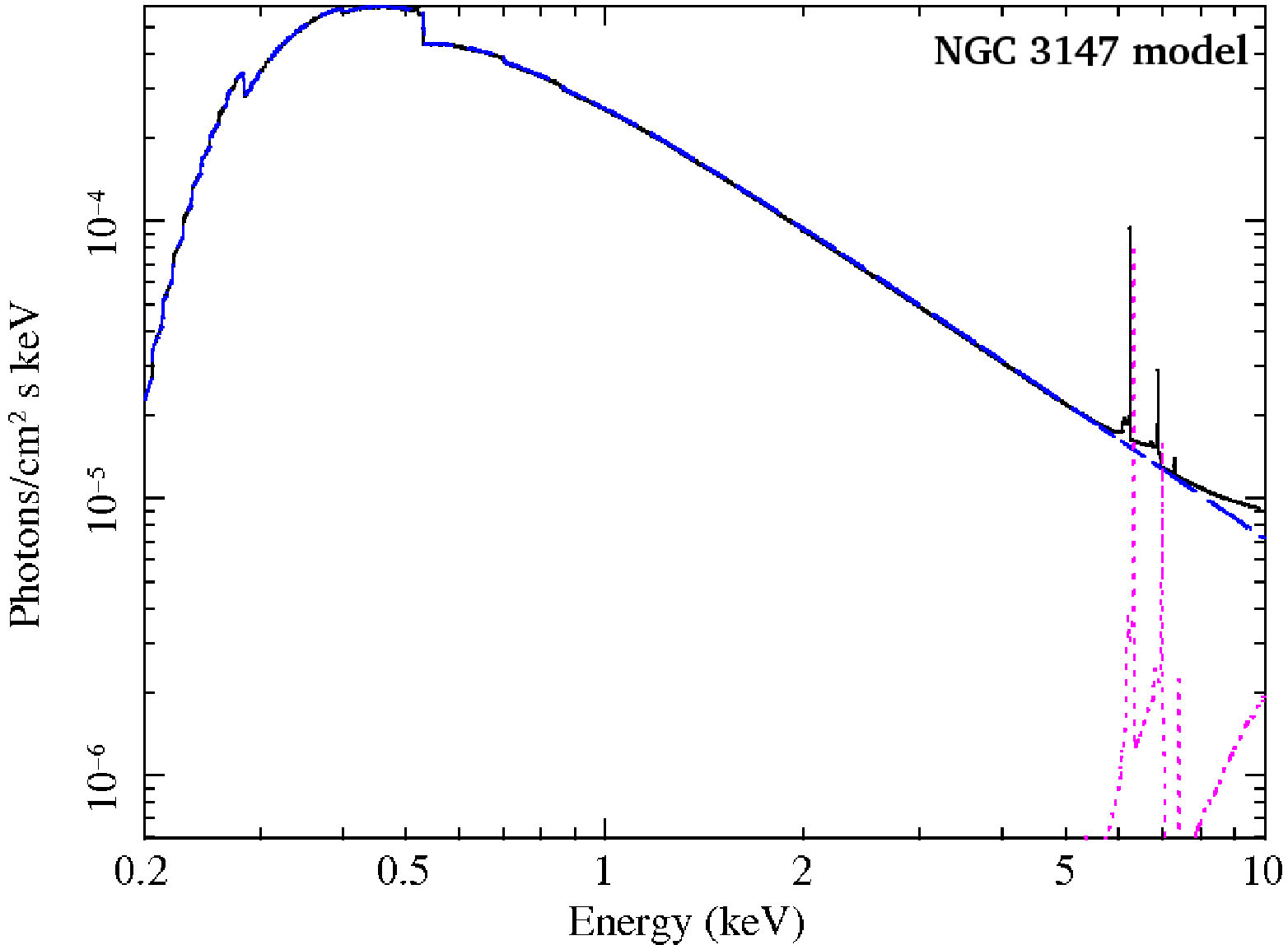}
 \includegraphics[width=80mm]{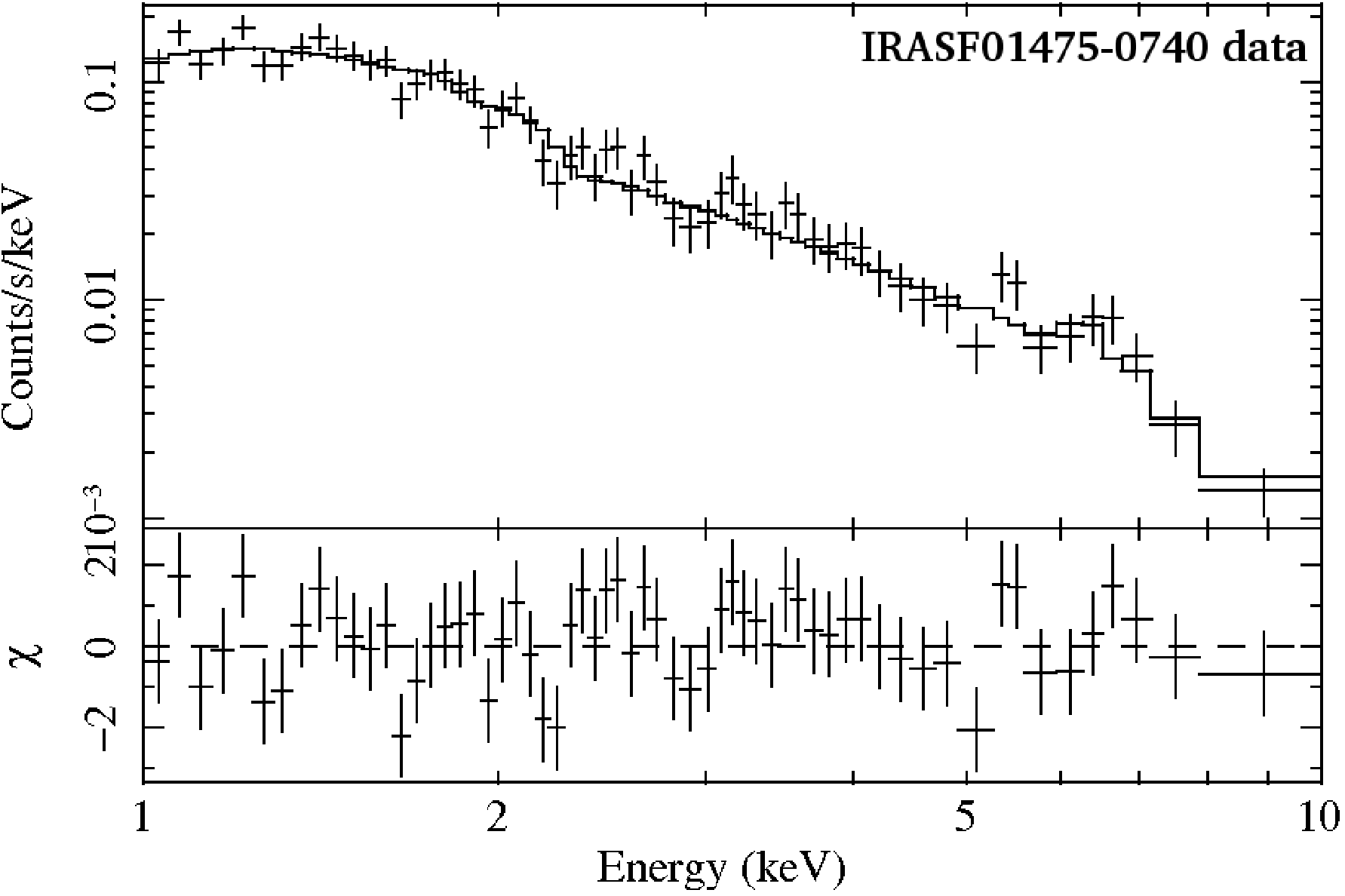}
 \includegraphics[width=80mm]{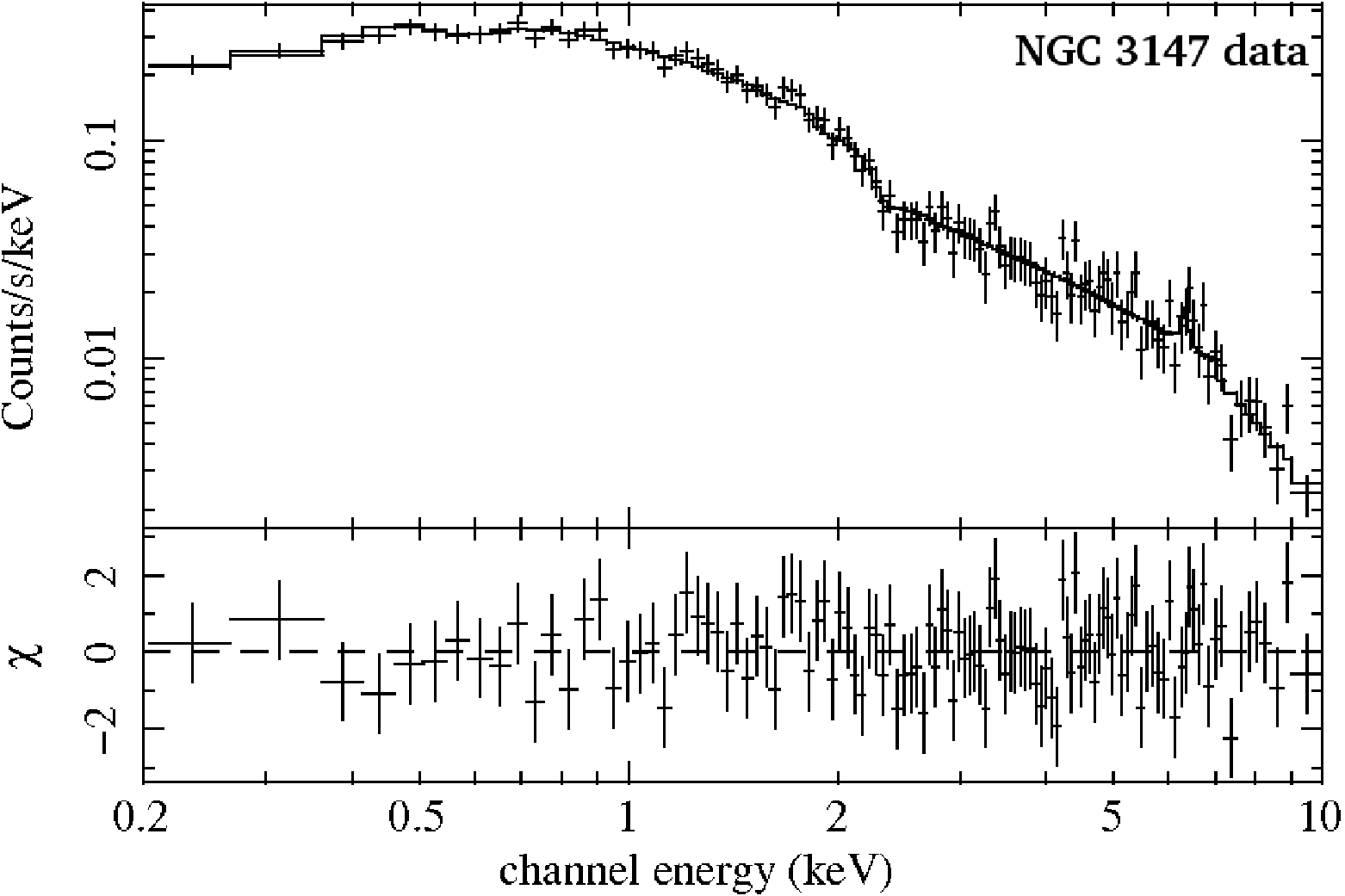}
 \includegraphics[width=80mm]{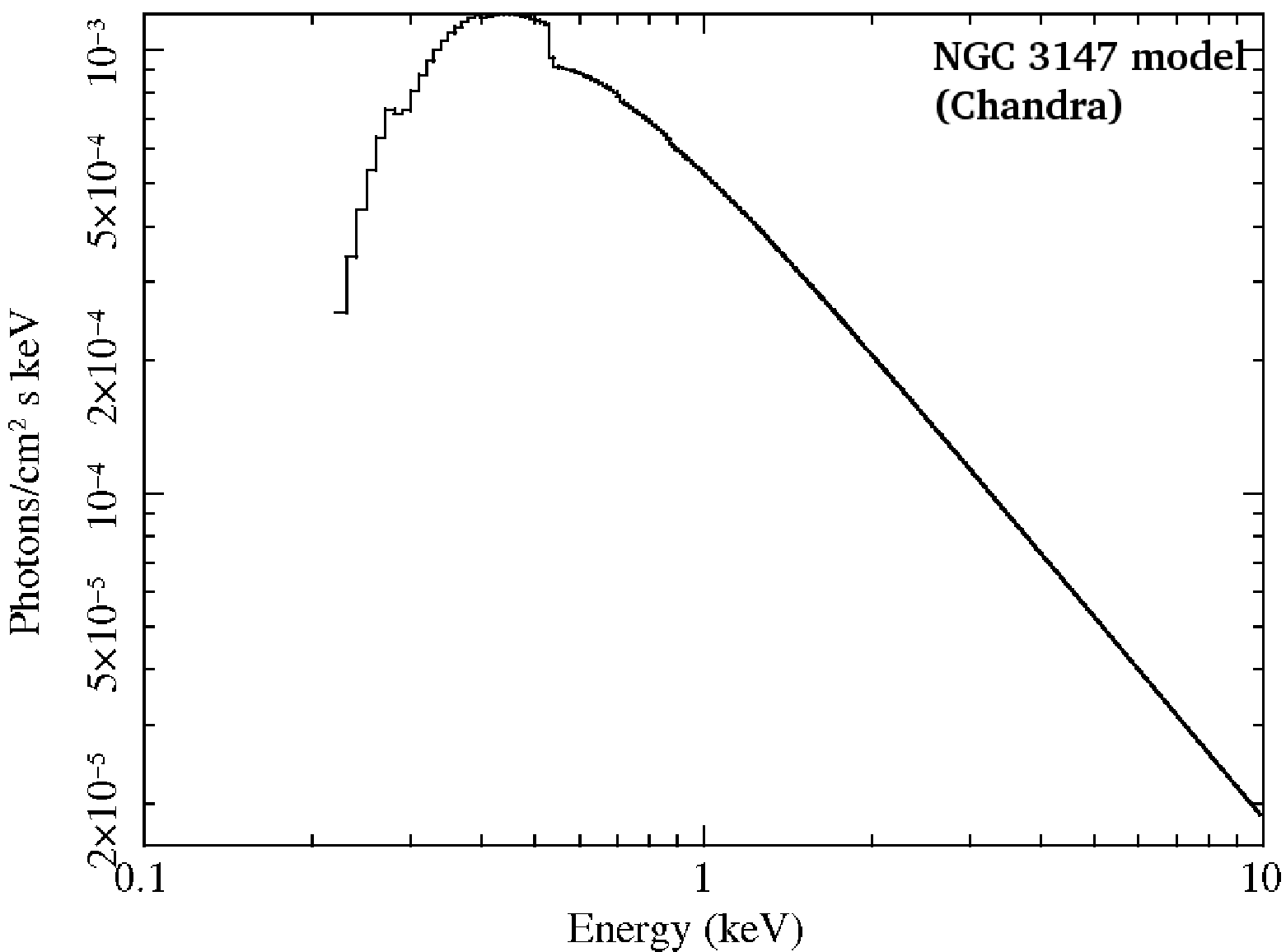}
 \includegraphics[width=80mm]{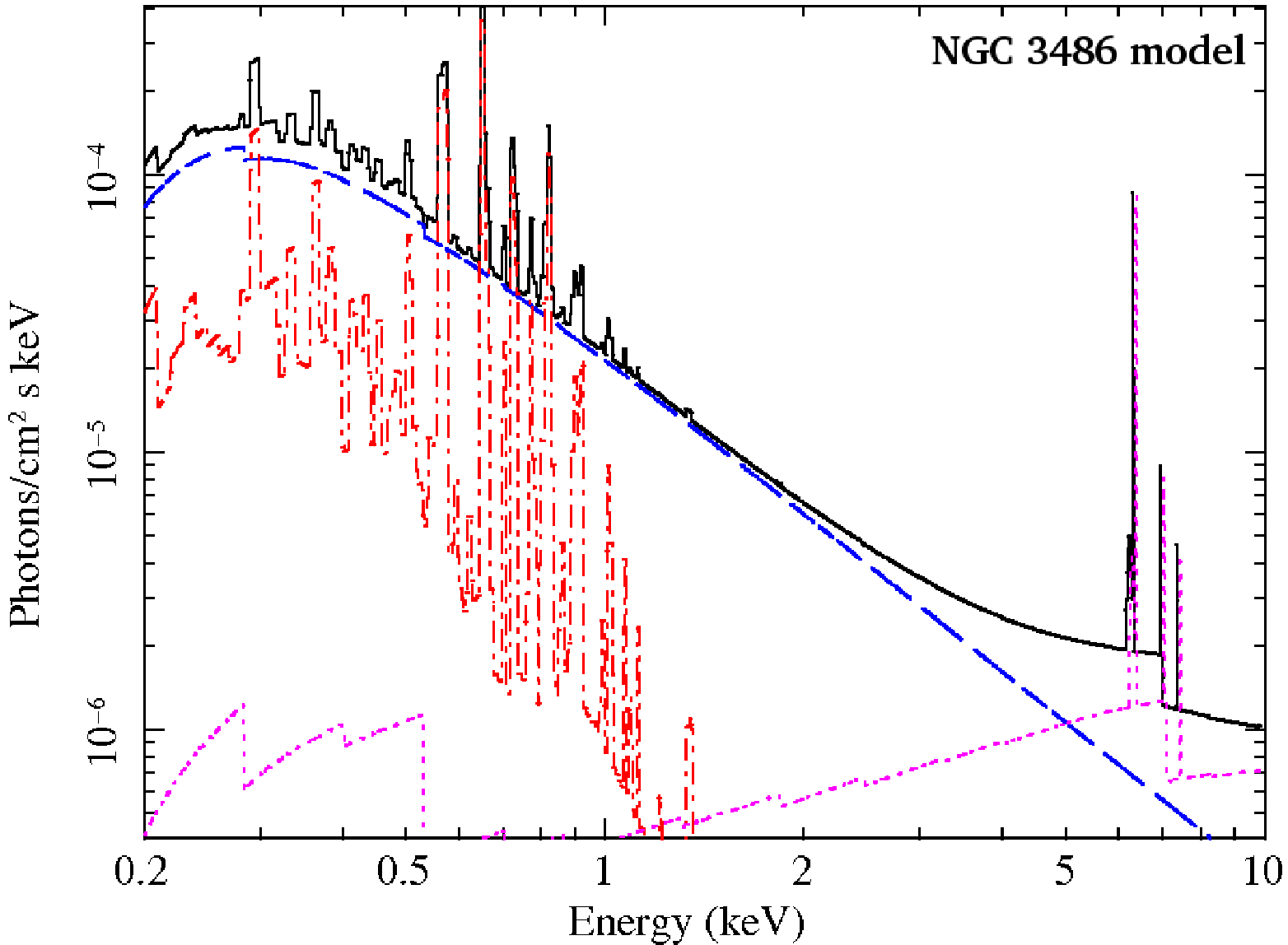}
 \includegraphics[width=80mm]{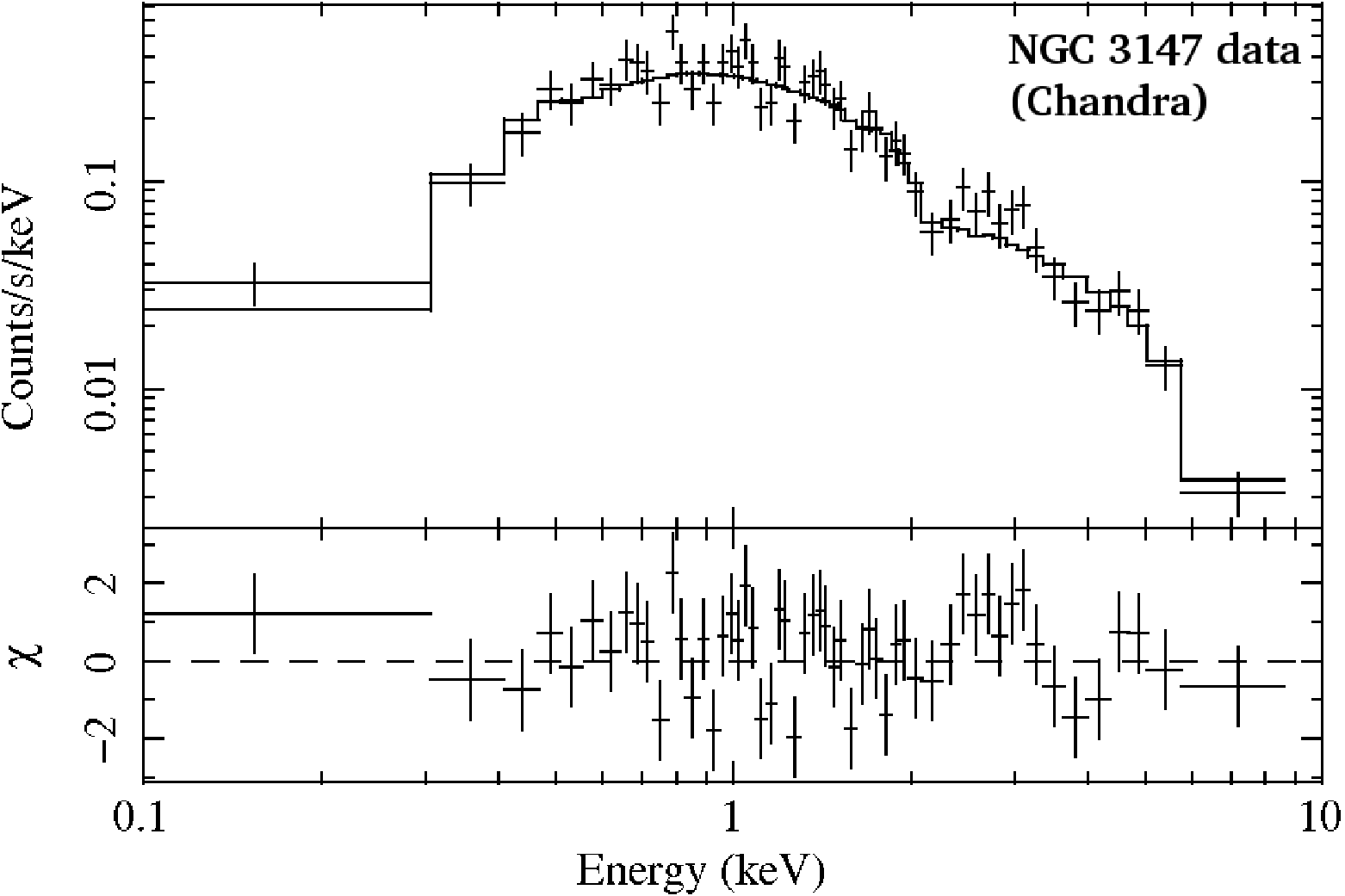}
 \includegraphics[width=80mm]{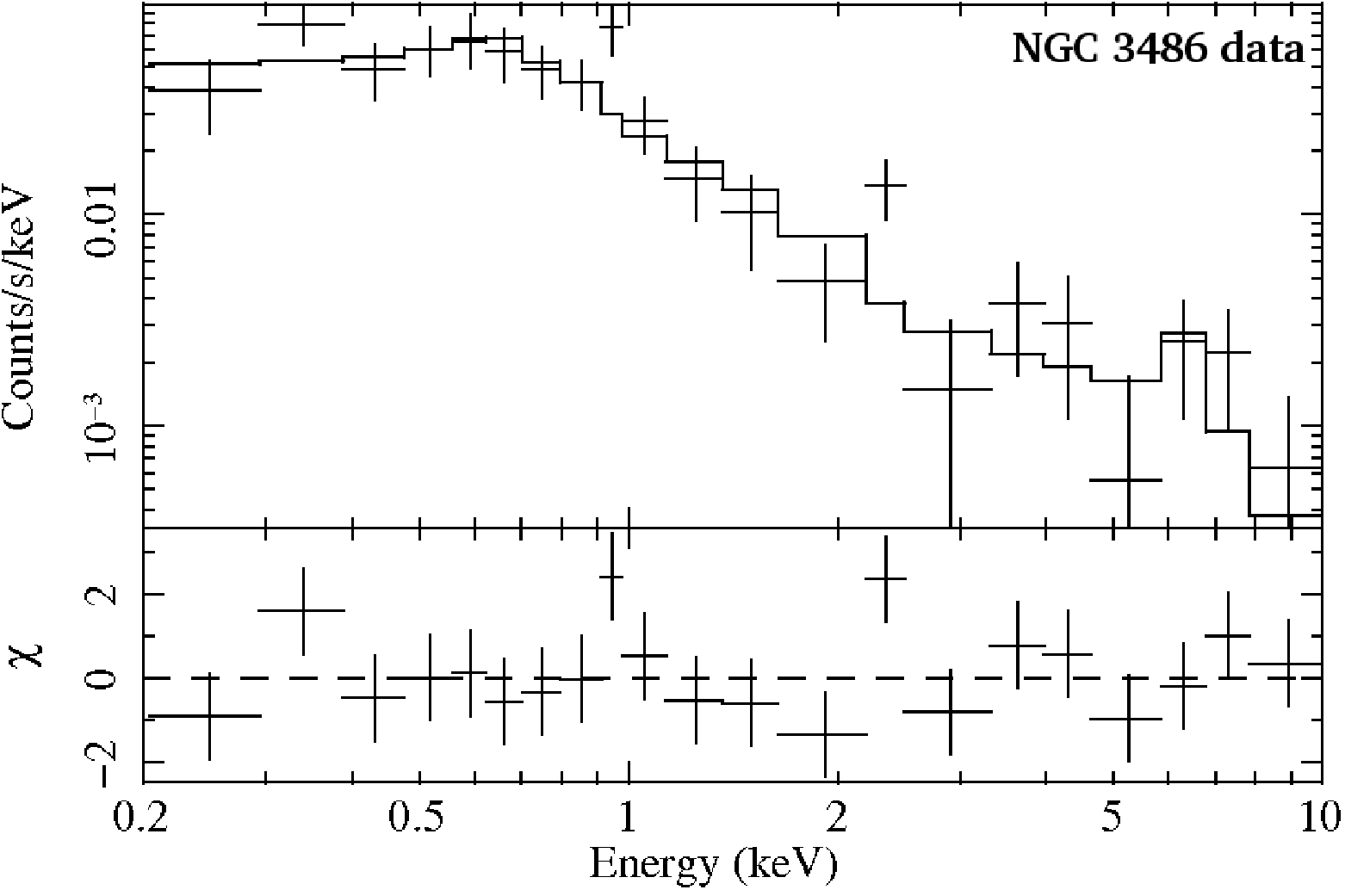}
 \end{minipage}
\end{figure*}

\begin{figure*}
\begin{minipage}{160mm}
 \includegraphics[width=80mm]{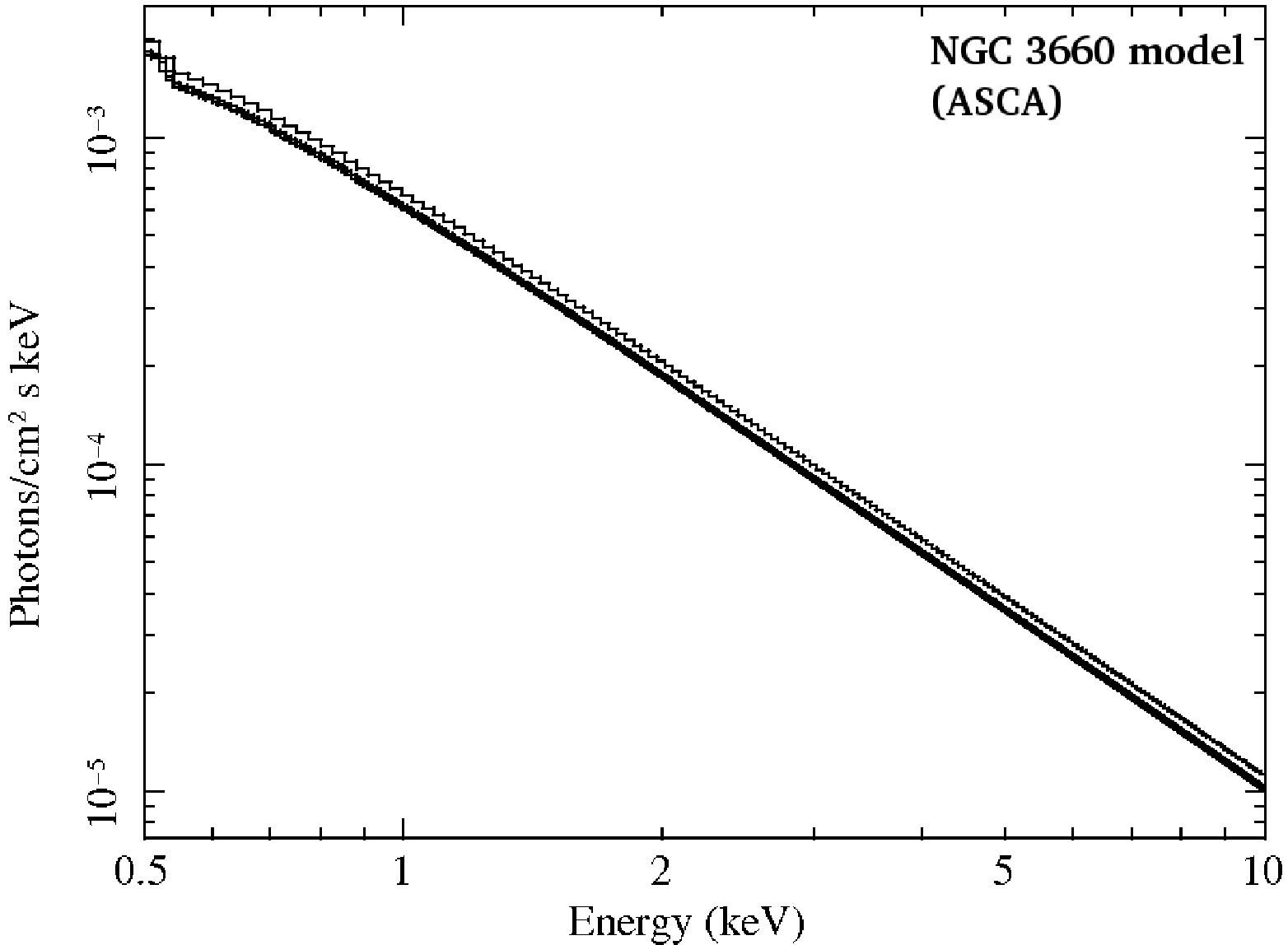}
 \includegraphics[width=80mm]{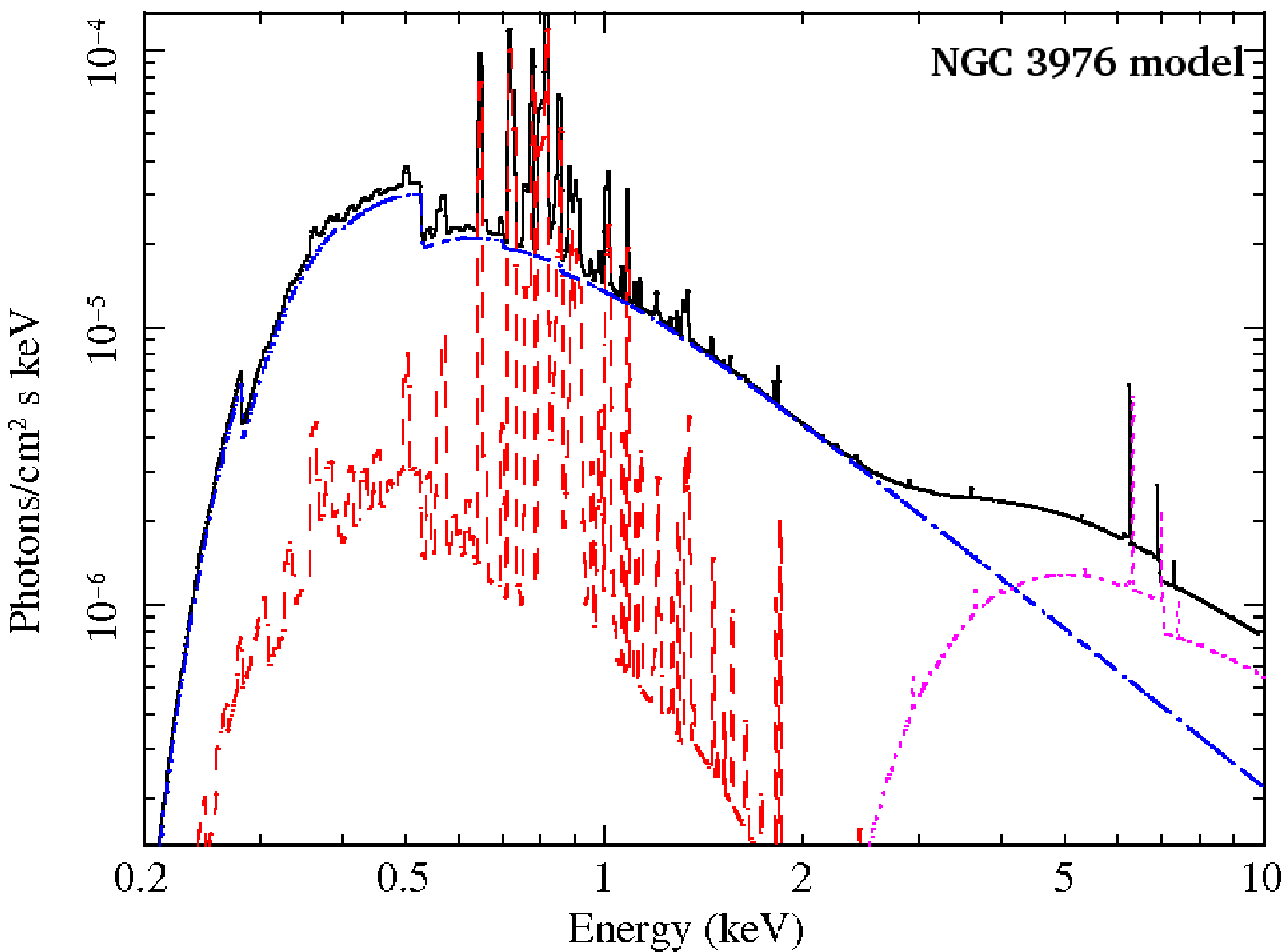}
 \includegraphics[width=80mm]{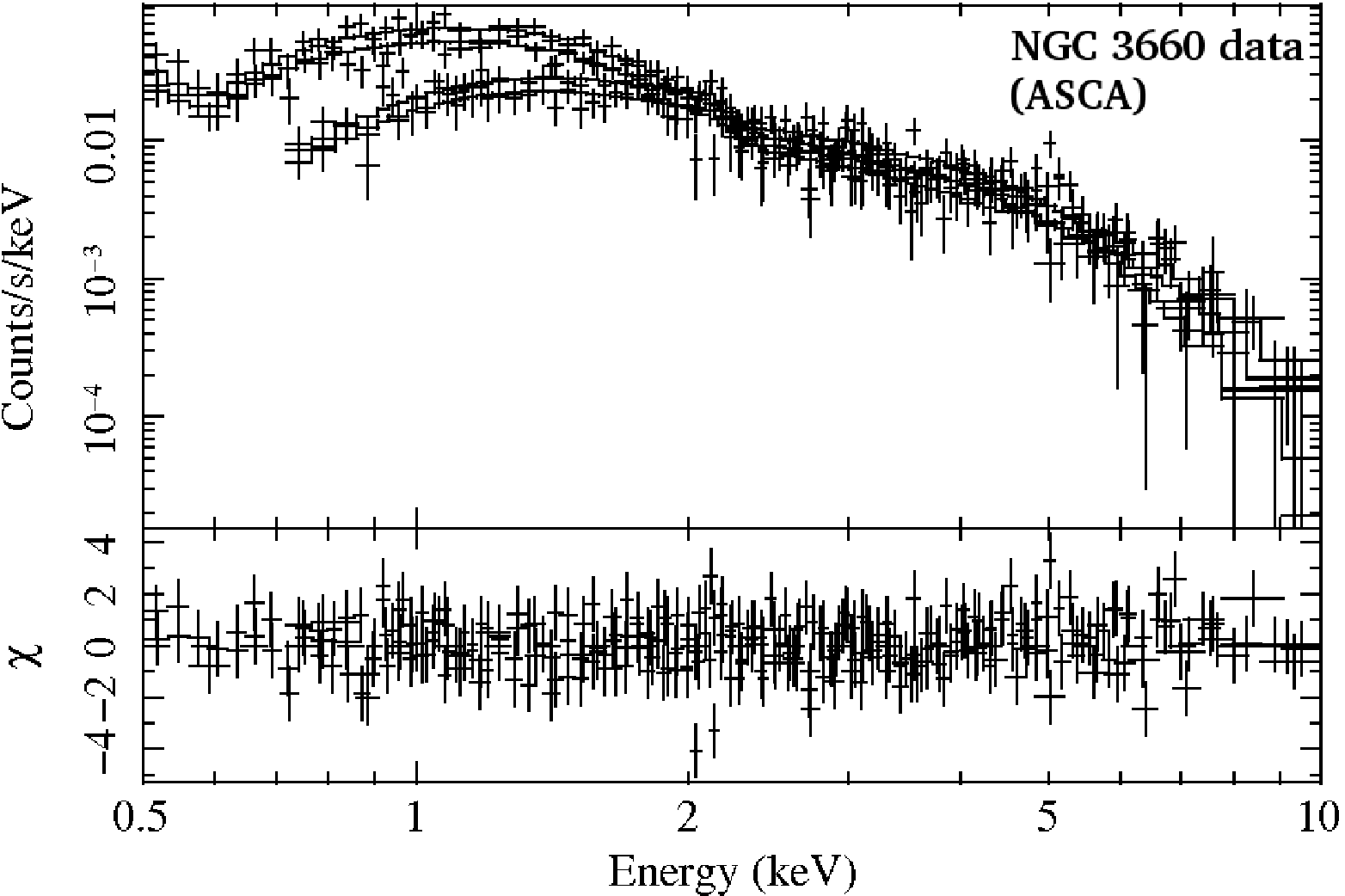}
 \includegraphics[width=80mm]{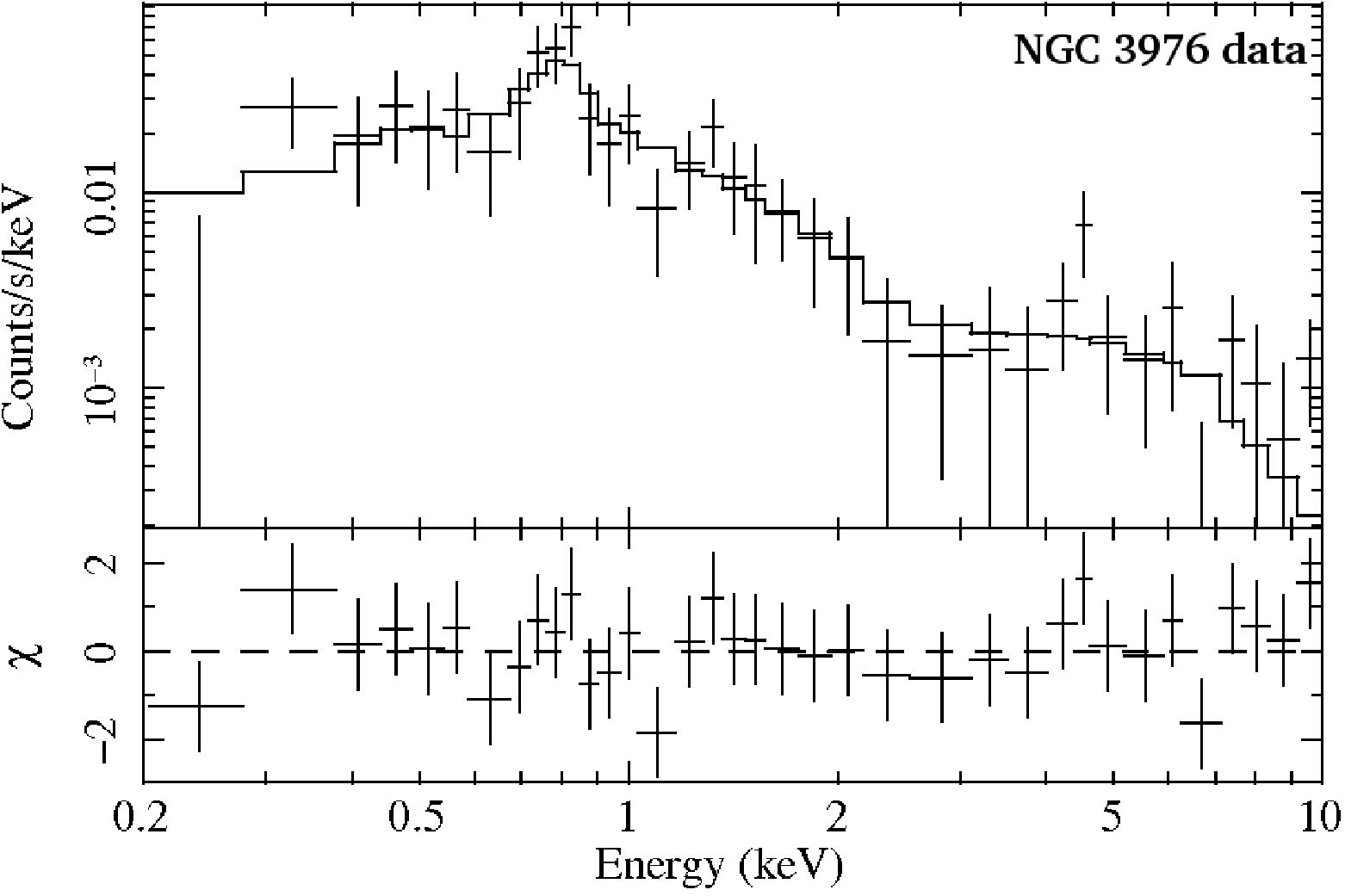}
 \includegraphics[width=80mm]{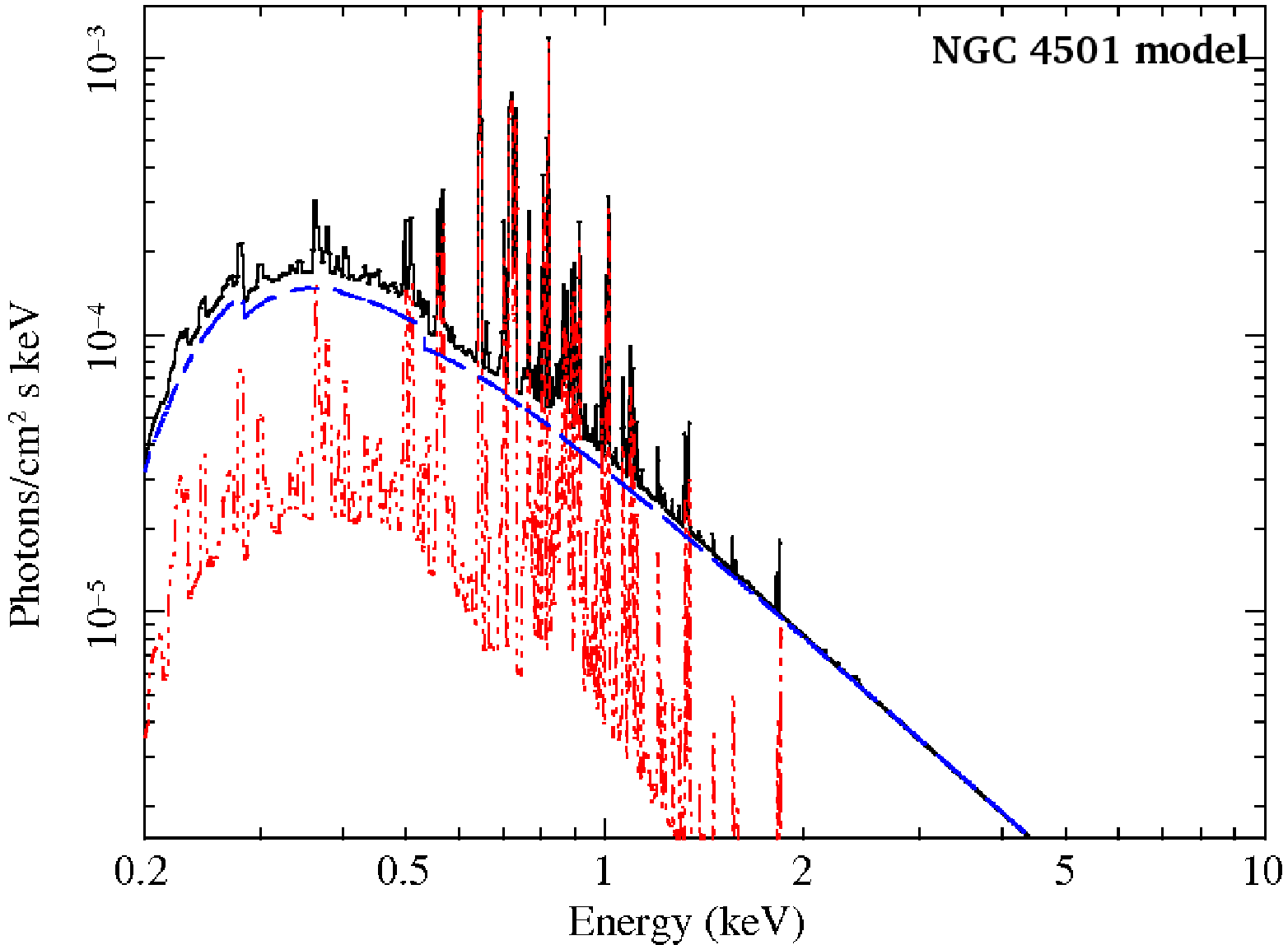}
 \includegraphics[width=80mm]{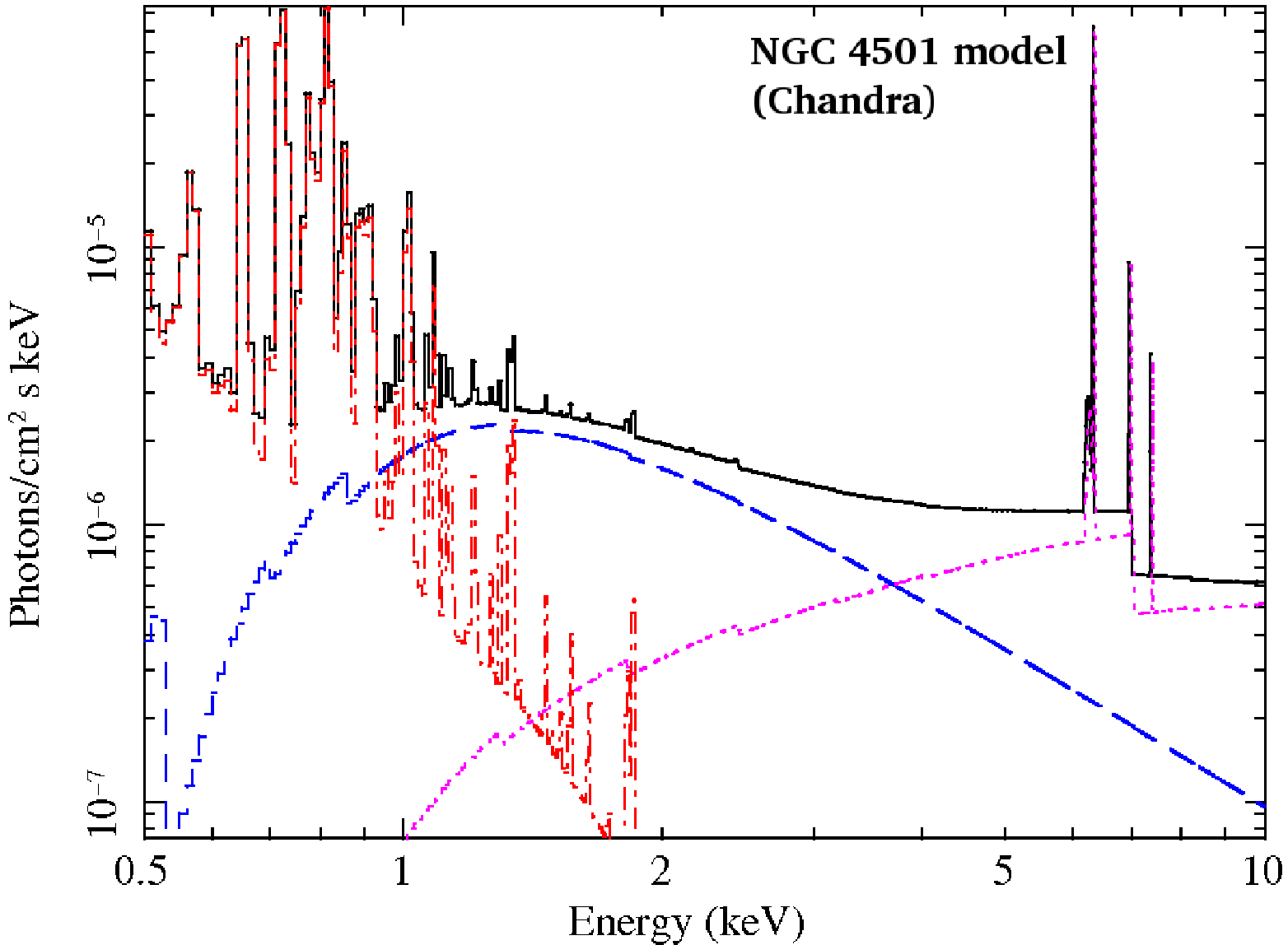}
 \includegraphics[width=80mm]{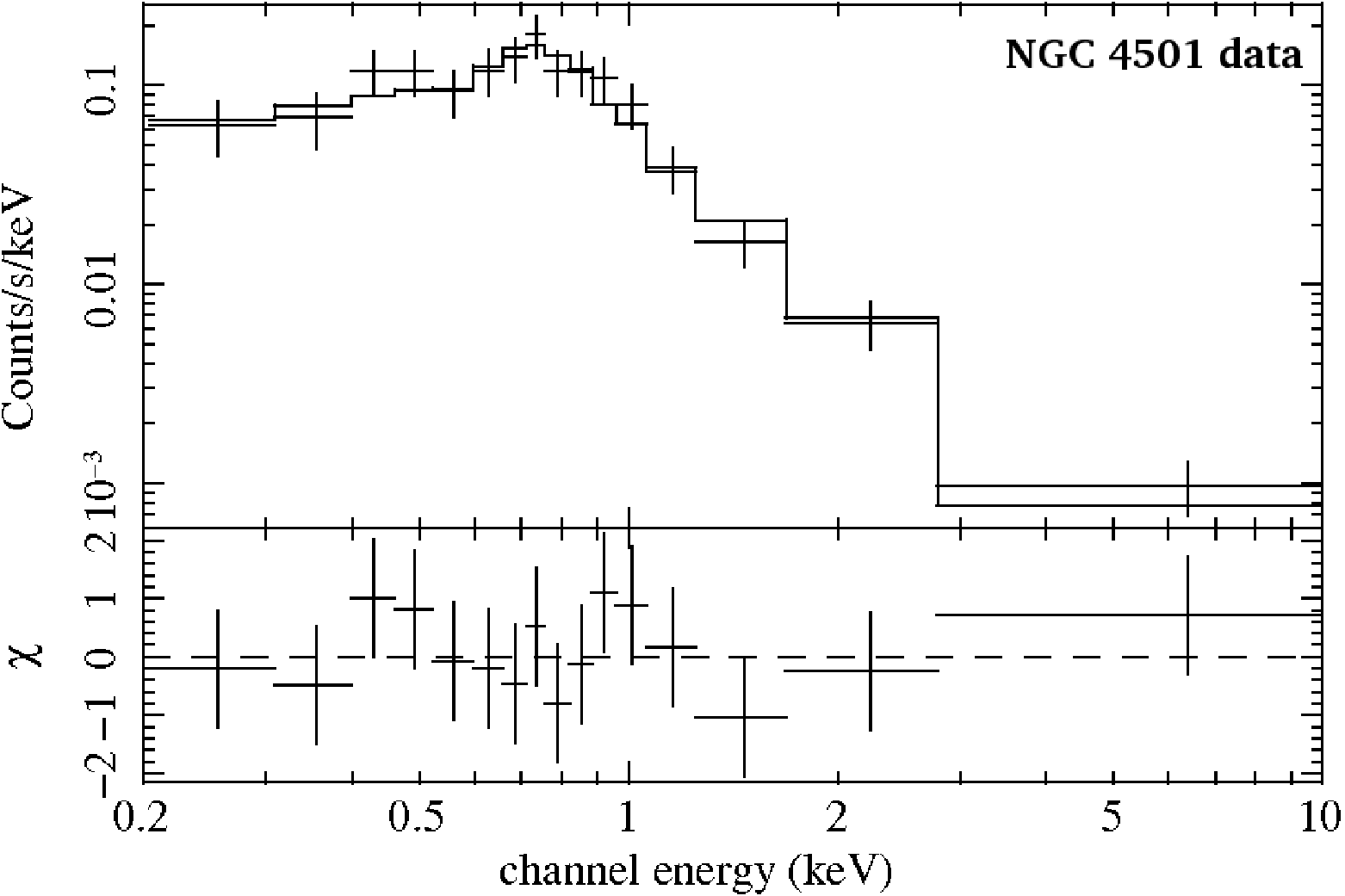}
 \includegraphics[width=80mm]{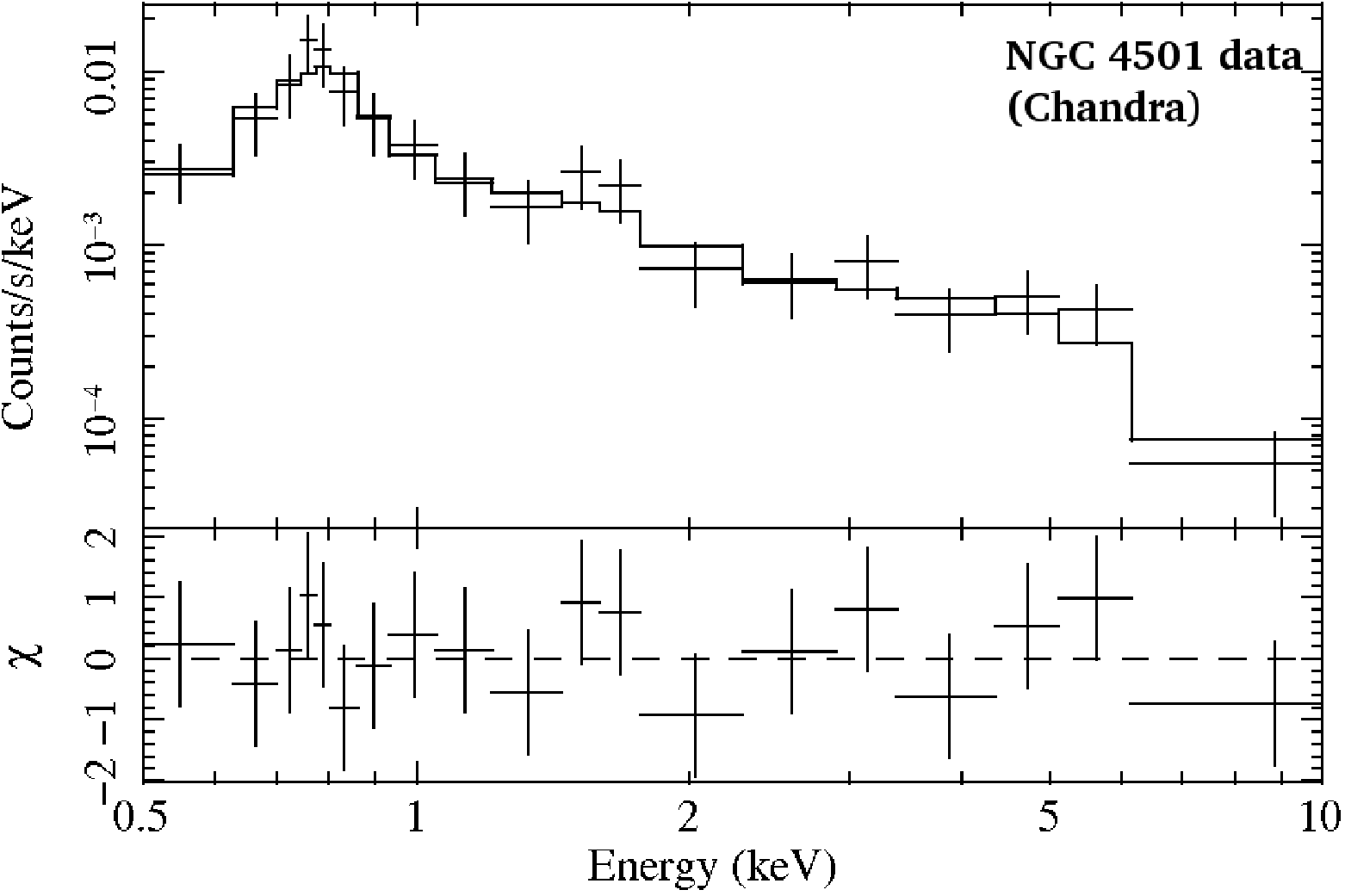}

 \caption{{\it XMM-Newton} (unless otherwise stated) spectra of IRASF 01475-0740, NGC 3147, NGC 3486, NGC 3660, NGC 3976 and NGC 4501, shown with best fit model from Table \ref{fitdat}.  In the plots of the best fit model, blue dashed lines represent the unabsorbed power-law, red dot-dashed lines represent the thermal component if present, and the magenta dotted lines represent the reflection or transmitted component if present.}
 \label{specfig}
\end{minipage}
\end{figure*}

\section{Discussion}

\subsection{Main Findings}

We have presented X-ray and multi-waveband data for a sample of bona fide Seyfert 2 galaxies, as defined by their optical line ratios, which appear unabsorbed in the X-ray. Our first key finding is that the 2-10 keV X-ray luminosities of the objects are low with respect to their mid-IR and optical [O\,{\sc iii}] luminosities. Under normal circumstances this would be interpreted straightforwardly as being due to suppression of the X-ray flux by an absorbing column, however a naive analysis of the X-ray spectra is at odds with this interpretation. It may be that the simplest interpretation of the spectra is misleading us: if scattered or host galaxy emission dominates the spectrum below $\sim 5$ keV, the spectrum may appear unabsorbed when in fact a harder nuclear components is present. We favour the latter interpretation in four objects. These show tentative evidence for a hard excess which is likely to be a hidden reflection or transmission spectrum indicative of a heavily obscured nucleus, accounting for the low hard X-ray luminosity of these objects. The most compelling case is NGC 4501, where the high resolution {\it Chandra} image reveals a hard source co-incident with the nucleus embedded in a number of softer emitting regions. The clear conclusion in this case is that is that  2-10 keV {\it XMM-Newton} data are  dominated by non-nuclear emission, accounting for the unabsorbed X-ray spectrum. We infer a similar conclusion for IRAS F01475-0740, NGC 3486 and NGC 3976. For two of our sources (NGC 3147 and NGC 3660) we see no signs of hidden reflection or transmission in the {\it XMM-Newton} X-ray spectra. NGC 3147 shows a single soft X-ray source co-incident with the nucleus in the Chandra image, and both exhibit X-ray variability. These observations clearly point to the idea that we are seeing the nuclei directly in these sources, so the evidence suggest that they might intrinsically lack a BLR. 

It therefore appears from our analysis that there are two populations of unabsorbed Seyfert 2 galaxies and that they appear unabsorbed in X-rays for different reasons entirely.  It may be that our view of the AGN is genuinely unobscured, and that the lack of broad lines in their optical spectra is due to an intrinsic absence or weakness of the BLR.  Alternatively though, it may be that the X-ray spectrum is misleading, and that the unabsorbed spectra are due to scattered or galactic soft X-rays which dominate the emission from a much more powerful obscured nucleus.  We discuss each in turn. 

\subsection{Genuinely unobscured Seyfert 2s}

\subsubsection{NGC 3147}
NGC 3147 has no hidden broad line region and a {\it Chandra} observations shows no discernible extra-nuclear sources.  Between observations by {\it Chandra} and {\it XMM-Newton}, the hard X-ray flux drops by a factor of $\sim 2$ indicating that it is variable in X-rays which means that we are likely to be seeing the nucleus of NGC 3147 directly. \cite{bianchi08} carried out simultaneous optical/X-ray observations of NGC 3147 to show that the apparent mismatch between optical and X-ray classification was not due to differential variability.  By noting a small equivalent width of the iron line and a large ratio between hard X-ray and [O {\sc iii}] fluxes, they come to the same conclusion that the nucleus of NGC 3147 is genuinely unobscured, and that this AGN must therefore intrinsically lack a broad line region.  We cannot however rule out the Compton thick nature of this source without observing it above 10 keV, where transmission from a Compton thick medium would dominate.

\subsubsection{NGC 3660}
The X-ray variability of this source on short time-scales and unabsorbed profile of its X-ray spectrum leads us to the conclusion that we are genuinely viewing the nucleus of NGC 3660 directly.  This leads us to believe that NGC 3660 also lacks a broad line region as we believe with NGC 3147. However, the origin of the X-ray variability is not necessarily nuclear as variability was recently discovered in the ultra-luminous X-ray (ULX) source, M82 X-1 \citep{mucciarelli06}.  A {\it Chandra} observation of NGC 3660 should be able to resolve any ULXs as we have shown for NGC 4501.  We also cannot rule out differential variability as the cause of the mismatch, as \cite{bianchi08} did for NGC 3147.  As has been observed in other X-ray variable Seyferts (e.g. NGC 4151, \cite{gaskell86}), the broad line flux has also been seen to vary.  If the broad line flux is observed in its low state, the object will be seen as type 2, despite being unobscured.  NGC 3660 will also need to be observed simultaneously in the optical and in the X-rays to rule out this possibility.

\subsection{Misidentified Heavily Obscured Seyfert 2s}

\subsubsection{IRASF 01475-0740}
\cite{bianchi08} use the small size of the iron line in NGC 3147 to argue against the Compton thick nature of that source.  However, IRASF01475-0740 also shows a small iron line, but in this object we know that heavy nuclear obscuration is occurring as shown by \cite{tran03}.  This spectropolarimetric study showed that IRASF01475-0740 has broad lines in its polarised optical spectrum, which are missing in its normal optical spectrum, confirming that our line of sight to the nucleus is blocked by optically thick material.  We show that a Compton thick source may exhibit only a moderate iron line if the continuum emission below 10 keV is dominated by a strong scattered continuum or extra-nuclear emission.  We demonstrated this by adding a second component to the fit of the X-ray spectrum, with a much larger column density than the first component,  by using the iron line as a constraint.  The intrinsic hard X-ray luminosity of this second component is a factor of $\sim 10$ greater than the observed luminosity.  

Assuming that the observed hard X-ray spectrum is indeed dominated by scattered light, we calculate an upper limit on the scattering fraction by fixing the column density of the second component to its lower limit.  The upper limit on the scattering fraction turns out to be $50\%$, far higher than the typical $\sim 3\%$ seen in Seyfert  2 galaxies \citep{cappi06}.  \cite{ueda07} observed a $<1\%$ scattered fraction in two Compton thick AGN, SWIFT J0601.98636 and SWIFT J0138.64001. They used that fraction to suggest that these objects have a large covering fraction of the torus, or a low abundance of the gas responsible for the scattering in those objects.  The inverse could be true for IRASF0175-0740 - it may have a low covering fraction of the torus, or a high abundance of the gas responsible for the scattering.  Scattered components are also often seen in reflection dominated sources (eg. NGC 1068; \citep{pier94}). If there is an underlying reflection component in IRASF0175-0740, this may also be dominated by scattered emission if the inclination angle of the torus is very high so that only a small part of the inner wall of the torus is directly visible and not self obscured.  This would give a very high scattered/reflected fraction as noted for model C in Table \ref{fitdat}.  

If the line emission seen here between 6 and 7 keV in fact originates from ionised iron, then we could be seeing reflection from ionised matter, as seen by \cite{nandra07_2} in IRAS 00182-7112.  The effect would be to increase the soft X-ray flux compared to neutral reflection as the lighter elements responsible for absorbing the soft X-rays will have been stripped of most of their electrons.  This would fit well with the soft spectrum seen in IRASF01475-0740.

The case of IRASF01745-0740 shows that Compton thick/intermediate AGN do not have to be reflection dominated with intense iron lines as they can be transmission dominated with a strong scattered component and no intense iron lines.

\subsubsection{NGC 4501}
By constraining hard X-ray emission to the nucleus in the {\it Chandra} observation of NGC 4501 we confirmed that this source is in fact Compton thick.  Since the extra-nuclear sources seen in the {\it Chandra} image lie outside the {\it XMM-Newton} spectral extraction region used, we surmise that the power-law and thermal emission which dominate over the reflection component in the {\it XMM-Newton} spectrum does not originate from ULXs or the AGN, but from diffuse galactic emission close to the nucleus.  If this is scattered emission from the nucleus, the scattered fraction is $\sim 3\%$ in this Seyfert.

\subsubsection{NGC 3486}
Fitting the hard excess in the X-ray spectrum of NGC 3486 with a reflection model yields an improved fit to this data, but the significance is too low to conclude that the source is Compton thick, as a reflection spectrum would imply. However, the {\it XMM-Newton} spectrum here, is in all ways similar to the {\it Chandra} spectrum of NGC 4501, which we showed to originate from a hard nucleus which is very likely Compton thick. As NGC 3486 is severely under-luminous in hard X-rays when compared to optical and mid-infrared luminosities, this conclusion seems likely for NGC 3486 too.  Again we need to observe this Seyfert 2 above 10 keV to confirm our hypothesis, or with the high angular resolution capabilities of {\it Chandra}.

\subsubsection{NGC 3976}
As with NGC 3486, the hard excess present in this X-ray spectrum is well fitted by the addition of a reflection component.  However, it is also well fitted by a transmission component.  Again, this spectrum is very similar to the {\it Chandra} spectrum of NGC 4501, so we conclude that this source is also heavily obscured.

\subsection{Wider Implications}

The discovery of AGN which do not have a broad line region has important consequences for the AGN unification model.  We can no longer exclusively invoke orientation based effects to explain the difference between type 1 and type 2 AGN.  There must be some other physical effect which means that the BLR is absent.  \cite{nicastro03} studied a sample of Seyfert 2s extracted from the spectropolarimetric study of \cite{tran03} and concluded that the presence of a hidden broad line region in Seyfert 2s is dependent on the rate at which matter is accreted by the central black hole.  They find that those which do not have a hidden broad line region have a mass accretion rate less than $10^{-3}$ Eddington units.  Thus, low accretion rate systems may not form a stable broad line region so they would not show broad lines in their spectra, yet will still be unobscured.

On the other hand, our multiwavelength analysis has shown a significant number (4 out of 6) Seyfert 2 galaxies in which the absorbing column appears to have been completely underestimated. The distribution of absorbing column densities is a key ingredient in models of the cosmic X-ray background, especially the fraction of Compton thick AGN, which is required to fit the 30 keV peak in the spectral energy distribution \citep{gilli07}. The derived $N_{\rm{H}}$ distribution from these models generally over-predicts the number of Compton thick AGN observed by many authors (e.g. \cite{risaliti99}) and also under-predicts the number of unobscured AGN.  Our work potentially provides an answer to this discrepancy, at least for the 12 ${\umu}$m sample.   From our small but well selected sample, we find that 2/3 of Seyfert 2 galaxies with an $N_{\rm{H}}$ reported to be $<10^{21}$ cm$^{-2}$ should in fact be designated with an $N_{\rm{H}}>10^{23}$ cm$^{-2}$, and in reality are probably Compton thick. The 12 ${\umu}$m sample contains 79 Seyfert 2s with good quality X-ray coverage, and of these 19 have been reported to be unabsorbed (24\%), compared to 4\% found by  \cite{risaliti99} for their sample of local Seyfert 2s. Combined with the consideration that many unabsorbed  Seyfert 2s of the 12 ${\umu}$m sample have a dubious classification as Seyfert 2, our findings alter the $N_{\rm{H}}$ distribution of the 12 ${\umu}$m sample to better agree with those derived from the X-ray background and other observations.  Misidentifying these heavily obscured objects due to contamination by the host galaxy, as may be the case in NGC 3486, 3976 and 4501, should be an effect only seen in local AGN, as in deep surveys, only the brightest AGN are selected, where the nucleus will dominate over any galactic emission, but if scattered nuclear emission dominates the soft X-ray then this could be a significant issue even at high redshift.

\section{Acknowledgements}
We thank the anonymous referee for their useful comments.  We thank the constructors and operators of {\it XMM-Newton}, {\it Chandra} \& {\it ASCA}.  MB would like to acknowledge the financial support from STFC.  This research has made use of the NASA/IPAC Extragalactic Database (NED) which is operated by the Jet Propulsion Laboratory, California Institute of Technology, under contract with the National Aeronautics and Space Administration. Funding for the SDSS and SDSS-II has been provided by the Alfred P. Sloan Foundation, the Participating Institutions, the National Science Foundation, the U.S. Department of Energy, the National Aeronautics and Space Administration, the Japanese Monbukagakusho, the Max Planck Society, and the Higher Education Funding Council for England. The SDSS Web Site is http://www.sdss.org/.  This research has made use of the Tartarus (Version 3.2) database, created by Paul O'Neill and Kirpal Nandra at Imperial College London, and Jane Turner at NASA/GSFC. Tartarus is supported by funding from PPARC, and NASA grants NAG5-7385 and NAG5-7067.

\bibliographystyle{mn2e}
\bibliography{bibdesk}

%\footnotetext\thanks{This paper has been typeset from a \TeX/\LaTeX file prepared by the author.}

\label{lastpage}
\end{document}